\documentclass[12pt]{iopart}
\usepackage{bm}
\usepackage{float}
\expandafter\let\csname equation*\endcsname\relax
\expandafter\let\csname endequation*\endcsname\relax
\usepackage{amsmath}
\usepackage{graphicx}% Include figure files
\usepackage{dcolumn}% equation table columns on decimal point
\usepackage{cite}
\usepackage{multirow}
\usepackage{footmisc}

\usepackage{soul}

\usepackage[dvipsnames]{xcolor}
\usepackage[colorlinks=true,allcolors=RoyalBlue]{hyperref}
\hypersetup{colorlinks=true}
\usepackage{amsfonts}
\usepackage[capitalize]{cleveref}
\usepackage{physics}

\DeclareMathOperator\erfc{Erfc}

\newcommand{\n}{\nonumber}

\newcommand{\rev}[1]{#1}

\usepackage{etoolbox}
\makeatletter
\newrobustcmd{\fixappendix}{%
  \patchcmd{\l@section}{1.5em}{7em}{}{}%
  \patchcmd{\l@subsection}{2.3em}{7em}{}{}%
}
\makeatother

\newcommand{\thetitle}{
Recoil of a driven tracer in a correlated medium
}

\begin{document}
\title[\thetitle]{\thetitle}
\author{Marcin Piotr Pruszczyk$^{1,2}$, Davide Venturelli$^{3,4}$ and
Andrea Gambassi$^{1,2}$}
\address{$^1$ SISSA --- International School for Advanced Studies, via Bonomea 265, 34136 Trieste, Italy}
\address{$^2$ INFN, via Bonomea 265, 34136 Trieste, Italy}
\address{$^3$ Laboratoire de Physique Th\'eorique de la Mati\`ere Condens\'ee, CNRS/Sorbonne Universit\'e, 4 Place Jussieu, 75005 Paris, France}
\address{$^4$ Laboratoire Jean Perrin, CNRS/Sorbonne Universit\'e, 4 Place Jussieu, 75005 Paris, France}
\ead{mpruszcz@sissa.it}
\begin{abstract}
 We study the stochastic dynamics of a Brownian
 particle after it is suddenly released from a harmonic trap moving with constant velocity through a fluctuating correlated medium, described by a scalar Gaussian field with relaxational dynamics and in contact with a thermal bath. 
 We show that, after the release,  
 the particle exhibits recoil, i.e., it moves in the direction opposite to the drag. As expected, this effect vanishes if the field equilibrates instantaneously. 
 The final value of the average position of the particle is reached algebraically in time in the case of conserved dynamics of the field or for non-conserved dynamics at the critical point. 
 Our predictions are expected to be relevant, at least qualitatively, to driven colloidal particles in liquid media close to critical points.  
\end{abstract}

\tableofcontents
\markboth{\thetitle}{\thetitle}

\section{Introduction}

Ever since the pioneering works of Einstein and Langevin~\cite{Einstein1906,Langevin_1908}, the dynamics of a mesoscopic particle immersed in a simple medium is commonly described in terms of the Langevin equation for the particle position $\mathbf{X}(t)$, featuring an instantaneous friction and a random force in the form of a Gaussian white noise. The validity of such a description hinges on the assumption of timescale separation, i.e., that the evolution of the medium occurs over timescales much shorter than those characterizing the motion of the particle. In contrast, complex media with %characterized by 
macroscopic relaxation times manifestly violate this assumption~\cite{Dhont_1996}.
Viscoelastic fluids~\cite{Larson_1999} may serve as an example: 
% of such media: 
indeed, storage and dissipation of energy within their complex microstructure typically result in macroscopically long stress-relaxation times. Consequently, dragging a colloidal particle through such a fluid drives the medium out of equilibrium, which in turn affects the statistics of the particle position~\cite{Dhont_1996}, as observed in active microrheology experiments~\cite{Squires_2005, Gazuz_2009, Gomez-Solano_2014, Gomez-Solano_2015, Jain_2021, Jain_2021_2}.

The %framework of the 
overdamped generalized Langevin equation (GLE~\cite{Mori_1965, Zwanzig_book}) offers a description of such a system on a coarse-grained scale by modeling the interaction between the medium and the particle via a friction %kernel 
$\int_{-\infty}^t  \mathrm{d}t'  \, \Gamma(t - t')\dot{\mathbf{X}}(t')$ --- characterized by the memory kernel $\Gamma(t)$ ---
together with a colored Gaussian noise whose temporal correlations ensure detailed balance. 
Such a description, though computationally convenient due to the linearity of the evolution equation, has its limitations. 
First, in general, the functional form of $\Gamma(t)$ is not known a priori: in practice, it is often assumed to be of the form $\Gamma(t) = \sum_ia_ie^{-t/\tau_i}$,
where the coefficients $\{a_i,\tau_i\}$ are obtained by fitting experimental or numerical data. \rev{For experiments involving viscoelastic media, such an analysis was performed e.g.~in Refs.~\cite{Caspers2023,Loos2024}.}
This simplified model --- in which each term of the sum may be heuristically visualized as due to a fictitious \textit{bath particle} harmonically coupled to the tracer --- is actually
able to capture the effective particle dynamics within the linear response regime.
Conversely, as the system is driven further out of equilibrium, it turns out that an increasing number of %more and more 
terms has to be included in the sum to fit accurately the data~\cite{Berner_2018,Elephant_2010}, possibly with $a_i<0$.
Second, in constructing the GLE from first principles (see, e.g., Ref.~\cite{Zwanzig_book}), the linear projection of the degrees of freedom of the medium that are being coarse grained may lead to a memory kernel that depends 
%in principle 
not only on the interaction between the particle and the medium, but also on other details of the system --- such as the strength of external (confining) forces acting solely on the particle~\cite{Daldrop_2017,Basu_2022}.

\begin{figure}
    \centering
  \includegraphics[width=0.85\linewidth]{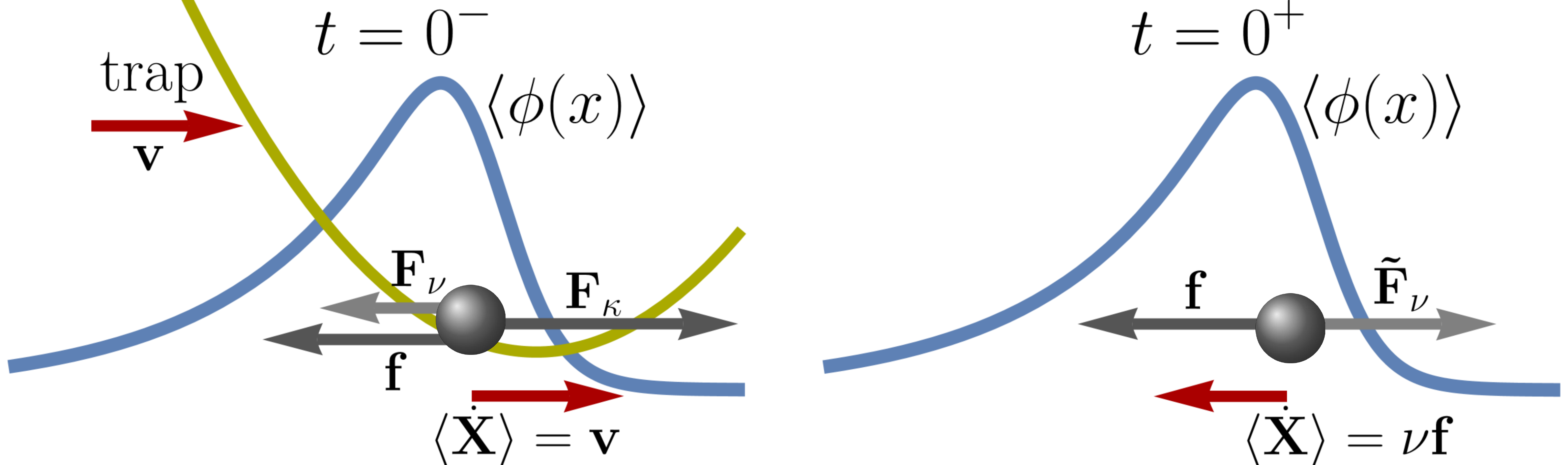}
    \put(-390,4){(a)}
    \put(-185,4){(b)}
\caption{Schematic representation of a Brownian particle (black sphere), driven by a moving optical trap (providing the potential represented by the dark yellow line) and  interacting with a fluctuating correlated field \rev{$\phi$} (dark blue line), \rev{which is preferably enhanced in the vicinity of the particle (see \cref{eq:Hint} with $\lambda>0$)}.
(a)~While the particle is driven, i.e., for $t<0$, the field attains an average configuration (blue line) which is stationary and spatially modulated in the reference frame co-moving with the trap. At time~$t = 0^-$, the particle is subject to the harmonic force $\mathbf{F}_{\kappa}$ exerted by the moving trap, to the field-induced force $\mathbf{f}$ defined in Eq.~\eqref{eq:force_def} \rev{which pulls the particle along the gradient of the field}, and to the friction force $\mathbf{F}_{\nu}$. 
(b)~At time $t = 0^+$, the particle is released from its confinement and thus it starts moving backwards (\textit{recoil}), due to the fact that $\mathbf{f}$ \rev{remains nonvanishing due to the slow relaxation of the field}, 
%\rev{\st{pulls it along the gradient of the field configuration  $\phi(\mathbf{x})$}}, 
and that the friction changes both its sign and amplitude $\mathbf{F}_{\nu} \mapsto \mathbf{\tilde{F}_{\nu}} = - \mathbf{f}$. 
}
\label{fig:cartoon}
\end{figure}

These considerations suggest the need for more sophisticated models which are able to describe the particle dynamics in complex media beyond the linear-response regime, including the peculiar phenomena exhibited by tracer particles in this context. One of them is the \textit{recoil} displayed by an optically confined particle dragged at constant velocity through a viscoelastic medium~\cite{Caspers2023,Gomez-Solano_2015, Robertson-Anderson_2018,Ginot_2022_rec, Cao_2023}:
after switching off %upon suddenly deactivating 
the confining force,
the particle turns out to move backwards, i.e., in the direction opposite to the previous drag, before coming to rest.
This somehow unexpected fact can actually be understood heuristically as follows. 
While the particle is driven through the medium, it locally perturbs its structure; accordingly,
the average position of the particle --- which reaches a steady-state value in the frame of reference co-moving with the trap --- is determined by the mechanical balance between 
the confining force due to the 
optical trap, the friction, 
and the overall force exerted on the particle by the nonhomogeneous structure acquired by the medium. This is schematically represented in \cref{fig:cartoon} \rev{(in which the field $\phi$ describes the relevant thermally fluctuating field within the medium, as clarified further below)}. 
%%
%%
%%
%%%%%%%%%%%%%%%%%%%%%%%%%%%%%%%%%%%%%%%%%%
%%%%%%%%%%%%%%%%%%%%%%%%%%%%%%%%%%%
%%
%%
%%
Generically, the latter force tends to increase the total friction, pulling the particle in the direction opposite to the drag.
When the trap is suddenly switched off, 
this force \textit{does not} vanish instantaneously: consequently,
the particle is temporarily pulled backward, 
until the medium has time to rearrange and relax to equilibrium. 

The recoil described above essentially stems from two key ingredients: (i)~a spatial structure acquired by the complex medium, and (ii)~a long relaxation time, which renders the effective particle dynamics non-Markovian.
A simple model 
that incorporates both these features
can be formulated in terms of the coupled overdamped dynamics of a particle and a scalar Gaussian field $\phi(\mathbf{x}, t)$, characterized by a macroscopic correlation length $\xi$ and a finite relaxation time~\cite{Demery_2010, Demery_2010_2, Demery_2011_Cas, Demery_2011, Dean_2011, Demery_2013, Gross_2021, Venturelli_2022, Basu_2022, Venturelli_2022_2parts, Venturelli_Gross_2022, Venturelli_2023, venturelli2023stochastic, Demery_Gabassi_2023,Muzzeddu_2025}. 
Modeling the interaction between the field and the particle via a coupling that is linear in $\phi$ allows one to integrate out the latter exactly,
yielding the effective (non-linear, non-Markovian) dynamics of the particle. 
This procedure connects the emergent memory terms that describe the motion of the particle with the underlying properties of the medium. 
Originally introduced to describe tracer diffusion in a thermally fluctuating environment~\cite{Demery_2011}, this minimal model has been more recently
interpreted as a simplified description of colloidal particles immersed in a 
medium close to criticality, such as a binary liquid mixture in the vicinity of the critical point of its demixing transition~\cite{Hertlein_2008, Gambassi_2009_exp, Paladugu_2016, Martinez_Entropy_2017, Magazzu_2019} --- with a particular interest in the emergence of fluctuation-induced  critical Casimir forces~\cite{Krech_book, Brankov_book ,Gambassi_2009,Gambassi2024-jb}.
%
%\ag{ADD HERE MY REVIEW IN SOFT MATTER, 2024... link in the file latex}
% https://pubs.rsc.org/en/content/articlelanding/2024/sm/d3sm01408h
%
In such media, characterized by 
extended
spatial and temporal correlations, $\phi$ should be regarded as the order parameter describing a second-order phase transition~\cite{Halperin_1977} (e.g., the relative concentration of the two components in the case of a binary liquid mixture). 
More broadly, such spatio-temporal correlations also affect the dynamics of inclusions in lipid membranes~\cite{Reister_2005, Reister_2010, Camley_2014}, microemulsions~\cite{Gompper_1994, Hennes_1996, Gonnella_1997}, or defects in ferromagnetic systems~\cite{Demery_2010, Demery_2010_2, Demery_2011_Cas, Demery_2011, Dean_2011, Demery_2013}.

A recent theoretical study~\cite{Venturelli_2023} investigated the motion of an overdamped particle driven through a correlated medium, within the framework described above. Notably, it was shown that the particle may exhibit oscillatory relaxation modes (generally forbidden in Markovian overdamped systems), analogous to those experimentally observed with a colloidal particle driven through a viscoelastic medium~\cite{Berner_2018}.
In order to further investigate the phenomenological similarity between viscoelastic and correlated media, in this work we examine the possibility of observing a correlation-induced analogue of the elastic recoil observed in experiments~\cite{Gomez-Solano_2015, Robertson-Anderson_2018,Ginot_2022_rec, Cao_2023}. 
To do so, in \cref{sec:Model} we consider a particle 
subject to a harmonic potential moving with a constant velocity $v$, coupled linearly to a scalar Gaussian field which follows a relaxational dynamics~\cite{Tauber_2014}. 
After the system attains a non-equilibrium steady state at long times in the reference frame co-moving with the trap, the harmonic trap is abruptly switched off, releasing the particle from its original confinement. 
The effective, nonlinear dynamics of the particle obtained by integrating out the field degrees of freedom is then solved perturbatively in the particle-field coupling in \cref{Sec:part_pos}. 
The subsequent average trajectory of the particle turns out to exhibit a recoil, which we characterize analytically and verify via numerical simulations of the system. Both the cases of locally conserved and non-conserved dynamics of the field are considered, also in the \textit{critical} limit in which the corresponding correlation length and relaxation time of the field diverge. We complement our study by the analysis of the steady-state field configuration presented in \cref{sec:NESS}, and finally in \cref{sec:conclusions} we summarize our results and present our outlook.

\section{The model}
\label{sec:Model}

We consider a system consisting of a tracer, modeling a colloidal particle, and a correlated medium, with the effective Hamiltonian~\cite{Venturelli_2022}
\begin{equation}
    \mathcal{H}[\phi,\mathbf{X}] = \mathcal{H}_\phi[\phi]  +
    \mathcal{H}_{\text{int}}[\phi,\mathbf{X}] + \mathcal{U}(\mathbf{X}, t) .
    \label{eq:hamiltonian_rec}
\end{equation} 
Here $\mathbf{X}=(X_1,X_2, \ldots, X_d)$ is the position of the particle in $d$ spatial dimensions, which is subject to a  time-dependent potential
\begin{equation}
    \mathcal{U}(\mathbf{X}, t)=  \frac{\kappa}{2}\left( \mathbf{X} - \mathbf{v}t \right)^2\vartheta(-t),
    \label{def:Uk_rec}
\end{equation}
describing a harmonic trap with stiffness $\kappa$. The trap has a center that moves with constant velocity $\mathbf{v}$, and it is switched off abruptly at $t=0$, as encoded by the Heaviside step function $\vartheta(-t)$.  
The correlated medium is modeled by a fluctuating scalar field $\phi(\mathbf{x}, t) \in \mathbb{R}$ characterized by the quadratic Hamiltonian~\cite{Halperin_1977}
\begin{equation}
        \mathcal{H}_\phi[\phi]= \int \mathrm{d}^d{\mathbf{x}}\left[ \frac{1}{2}(\nabla\phi)^2+\frac{1}{2}r\phi^2\right].
        \label{eq:gaussian_hamiltonian_rec}
\end{equation}
The parameter $r>0$ determines the spatial extension of correlations of the field, and is related to the correlation length $\xi$ via $\xi= r^{-1/2}$. In this respect, the thermally fluctuating medium undergoes a second-order phase transition at its critical point $r = 0$, characterized by a diverging correlation length $\xi$ and relaxation time.
Additionally, the particle is coupled linearly to the field in \cref{eq:hamiltonian_rec} by
\begin{equation}
    \mathcal{H}_\text{int}[\phi,\mathbf{X}] = -\lambda \int \mathrm{d}^d  \mathbf{x}\,\phi(\mathbf{x})V(\mathbf{x}-\mathbf{X}).
    \label{eq:Hint}
\end{equation}
In passing, we note that such a coupling is invariant under translations in space. The interaction kernel $V(\mathbf{x})$ actually models the shape of the particle and is normalized by requiring that its integral over the whole space is one.  
Accordingly, the strength of the interaction between the field and the tracer is determined solely by the value of the parameter $\lambda$. 
Note that the coupling \eqref{eq:Hint} breaks the $\phi \mapsto - \phi$ symmetry of $\mathcal{H}_\phi[\phi]$
and thus, for positive $\lambda V(\mathbf{x})$, it favors configurations of the field in which $\phi$ is enhanced in the vicinity of the particle\footnote{\rev{Recall that $\phi$ can be interpreted,
for instance, as
%as the order parameter of the second order phase transition, e.g, 
the relative concentration of the two components of a binary liquid mixture close to its demixing point. In this case, the enhancement of the field around the particle 
%can be interpreted as  
may indicate preferential adsorption of one of the two components by the surface of the particle.
%on the ``surface of the particle", as it is done in the discussion of antisymmetric boundary conditions in the context of critical Casimir forces \cite{Gambassi2024-jb}.
}}. 
Additionally, $V(\mathbf{x})$ is assumed to be spherically symmetric and characterized by a single lengthscale, i.e.,  $V(\mathbf{x}) =   \hat{V}(|\mathbf{x}|/R)$, where $R$ plays the role of the effective radius of the particle. 
The Gaussian kernel
\begin{equation}
    V(\mathbf{x}) = \frac{1}{(2 \pi R^2)^{d/2}}\exp\left(-\frac{\mathbf{x^2}}{2 R^2}\right)
    \label{eq:Gauss_kernel}
\end{equation}
may serve as an example. 
In this effective description of the interaction between the particle and the field introduced in Eq.~\eqref{eq:Hint}, 
the ``particle'' is identified with the interaction kernel $V(\mathbf{x})$, and the 
field
has non-zero values in its interior. 
Hence, the model introduced here characterizes the coupling of the particle to the field in terms of two effective parameters, namely the strength $\lambda$ and range $R$ of the interaction.

\subsection{Dynamics}

The dynamics of the particle is described by the overdamped Langevin equation 
\begin{equation}
        \dot{\mathbf{X}}= -\nu \bm{\nabla}_{\mathbf{X}} \mathcal{H}   + \bm{\xi} ,\label{eq:particle_rec}
\end{equation}
where $\nu$ is the mobility of the particle, and the stochastic noise $\bm{\xi}$ is a Gaussian variable with vanishing mean and variance given by (for simplicity the Boltzmann constant is set to unity here and henceforth, i.e., $k_B \equiv 1$)
\begin{equation}
    \left \langle\xi_{i}(t) \xi_{j}(t') \right \rangle = 2\nu T \delta_{ij}\delta(t-t'),
    \label{eq:part_noise}
\end{equation}
where $T$ is the temperature of the equilibrium heat bath within which the particle moves. 
Note that, assuming $\lambda V(\mathbf{x})>0$, the force $\mathbf{f}\left[ \phi,\mathbf{X} \right]$ exerted on the particle due to the coupling~\eqref{eq:Hint}, given by 
\begin{equation}
    \mathbf{f}\left[ \phi,\mathbf{X} \right] = - \bm{\nabla}_{\mathbf{X}}\mathcal{H}_{\mathrm{int}}[\phi, X] = \lambda \int \mathrm{d}^d\mathbf{x} \, V(\mathbf{x} - \mathbf{X}) \bm{\nabla}_{\mathbf{x}}\phi(\mathbf{x}),
    \label{eq:force_def}
\end{equation}
pulls the particle along the gradient of the field. For future convenience, we note that Eq.~\eqref{eq:force_def}, and consequently the
dynamics of the particle, exhibit invariance under $\{\phi, \lambda \} \mapsto \{-\phi, -\lambda \}$.
The evolution of the 
field
$\phi$ is governed by the relaxational dynamics~\cite{Tauber_2014}
\begin{equation}
\partial_t\phi(\mathbf{x},t)= -D_{\alpha} (i \nabla)^\alpha
\frac{\delta{\mathcal{H}(\phi, \mathbf{X})}}{\delta \phi(\mathbf{x},t)} + \eta(\mathbf{x},t),
\label{eq:field_rec} 
\end{equation}
where $D_{\alpha}$ is the mobility of the field\rev{, and $\alpha \in \{ 0, 2\}$}. Here, for $\alpha = 0$ the dynamics of the field does not obey any local conservation, while for $\alpha = 2$, the field is locally conserved by the dynamics.  In fact, Eq.~\eqref{eq:field_rec} can then be cast in the form $\partial_t \phi = - \bm{\nabla} \cdot \mathbf{J}(\mathbf{x}, t)$, where $\mathbf{J}$ is the corresponding current. We note that the cases $\alpha =0$ and $\alpha = 2$ correspond to the so-called model A and model~B, respectively, according to the classification of Ref.~\cite{Halperin_1977}. 
In the current analysis, however, we neglect the self-interaction term $\propto \phi^4$ present in the original formulation of these models, and study them within their Gaussian approximation. Note that the physical dimension $[D_{\alpha}]$ of the field mobility $D_\alpha$ depends on the choice of the dynamics. In particular, $[D_{\alpha}] = \mathcal{L}^{z}\mathcal{T}^{-1}$, where $z=  2+\alpha$ is the dynamical critical exponent of the field, $\mathcal{L}$ is a unit of length, and $\mathcal{T}$ is a unit of time.

Both the field and the particle are assumed to be in contact with the same heat reservoir characterized by the temperature $T$. Consequently, the noise 
$\eta(\mathbf{x}, t)$  acting on the field $\phi(\mathbf{x}, t)$  is a Gaussian variable with vanishing mean and characterized by the variance
\begin{equation}
    \left \langle\eta(\mathbf{x},t)\eta(\mathbf{x}',t')\right \rangle= 2D_{\alpha}T (i \nabla)^\alpha \delta^{(d)}(\mathbf{x}-\mathbf{x}')\delta(t-t').
    \label{eq:field_noise_rec}
\end{equation}
The correlations of the noise given by Eqs.~\eqref{eq:part_noise} and \eqref{eq:field_noise_rec} satisfy the fluctuation-dissipation relation. 
Accordingly, in the absence of external driving, i.e., for $\mathbf{v} = 0$, the joint dynamics of the particle and the field is invariant under time-reversal, and the joint probability distribution function of the position of the particle and the field configuration in the equilibrium steady state is given by the Boltzmann distribution $\propto \exp\left({-\mathcal{H}[\phi, \mathbf{X}]}/T\right)$, where $\mathcal{H}$ is the Hamiltonian in Eq.~\eqref{eq:hamiltonian_rec}.

In order to solve for the dynamics of the field 
as a function of the particle's position,
it will prove useful to express the field in terms of its Fourier transform as
\begin{equation}
    \phi(\mathbf{x,t}) = \int \frac{\mathrm{d}^d  \mathbf{q}}{(2\pi)^d}e^{i \mathbf{q}\cdot \mathbf{x}}\phi_{\mathbf{q}}(t), \qquad \phi_{\mathbf{q}}(t) = \int \mathrm{d}^d  \mathbf{x}\,e^{-i \mathbf{q}\cdot \mathbf{x}} \phi({\mathbf{x}},t),
    \label{eq:FT}
\end{equation}
where the wavevector $\mathbf{q}$ labels each mode of the field. In these terms, 
the dynamics of the field imposed by Eqs.~\eqref{eq:hamiltonian_rec}, \eqref{eq:gaussian_hamiltonian_rec}, \eqref{eq:Hint}, \eqref{eq:field_rec}, and \eqref{eq:field_noise_rec}
takes the form
\begin{gather}
    \left[ \partial_t + D_{\alpha}q^{\alpha}(q^2 + r)\right]\phi_{\mathbf{q}}(t) = \lambda D_{\alpha}q^{\alpha}V_{\mathbf{q}}e^{-i\mathbf{q}\cdot \mathbf{X}(t)}  + \eta_{\mathbf{q}}(t), 
    \label{eq:field_four}
    \\[2mm]
    \left \langle \eta_{\mathbf{q}}(t)\eta_{\mathbf{q'}}(t')  \right\rangle = 2D_{\alpha}q^{\alpha} T\delta(t - t') (2 \pi)^d\delta^{(d)}(\mathbf{q} + 
    \mathbf{q'}).
    \label{eq:noise_corr_ft}
\end{gather}
Here, $V_{\mathbf{q}}$ is the Fourier transform (see \cref{eq:FT}) of the interaction kernel introduced in \cref{eq:Hint}.
Similarly, the dynamics of the particle position
can be expressed as\footnote{
%\ag{CHECK, I MODIFIED THIS:} 
% \dav{Nope, it must be exactly $\lambda$ (and not generically $h$) in order to counteract this growth...}\ag{ahah! then it should be $\lambda V_{q=0}$ but $V_{q=0}=1$, right?}\dav{Right! I'll add it for more clarity}
Note that, in the case of model~A ($\alpha=0$), the average field mode $\phi_{\bm{q}=\bm{0}}$ may take arbitrarily large absolute values as criticality is approached for $r\to 0$. However, this fact is inconsequential for the particle dynamics: indeed, the mode with $\vb{q}=\bm{0}$ does not contribute to \cref{eq:part}. In order to counteract this growth, one would need to add a suitable chemical potential --- e.g., by replacing $\mathcal H_\phi \mapsto \mathcal H_\phi + 
%h  
\lambda V_{\vb q=\bm 0}
\int \dd{\bm{x}} \phi(\bm{x}) $ in \cref{eq:gaussian_hamiltonian_rec}. 
}
\begin{equation}
         \dot{\mathbf{X}}(t) = -\nu \kappa  \left( \mathbf{X}(t) - \mathbf{v}t\right)\vartheta(-t) + i\lambda\nu \int \frac{\mathrm{d}^d  \mathbf{q}}{(2\pi)^d}\mathbf{q}V_{-\mathbf{q}}e^{i\mathbf{q}\cdot \mathbf{X}(t)}\phi_{\mathbf{q}}(t) + \bm{\xi}(t). 
         %\qquad t<0. 
         \label{eq:part}
\end{equation} 
Note that for $\lambda = 0$ and $t<0$, according to Eqs.~\eqref{eq:field_four} and \eqref{eq:part}, each component of the particle position $\mathbf{X}$ and each mode of the field $\phi_{\mathbf{q}}$ are described by  non-interacting Ornstein-Uhlenbeck processes with relaxation times given, respectively, by
\begin{align}
    \tau_X^{-1} &= \nu \kappa, 
    \label{eq:tau_xd}
    \\ 
    \tau_{\phi}^{-1}(q) &= \alpha_q \equiv D_{\alpha}q^{\alpha}(q^2 + r). 
    \label{eq:alpha_q}
\end{align}
Accordingly, the field $\phi(\mathbf{x},t)$ exhibits not only long-range spatial correlations across the distance set by $\xi$, but also correlations over long times $\propto \xi^z$. In particular, for model~A, the relaxation time of the 
long-wavelength (small-wavevector) modes diverges as the system reaches criticality ($r \to 0$). Such a divergence is present in model B both in the critical and off-critical case (i.e., finite $r$) due to the local conservation of the order parameter~\cite{Tauber_2014}. Consequently, the coupling between the tracer and the slow-relaxing field yields a non-Markovian effective dynamics of the particle, due to a generic lack of separation of timescales.

Note that, due to the simplicity of its formulation, this minimal model may not only describe a near-critical medium, but also serve as a simple approximation of various complex media exhibiting both spatial and temporal correlations.

\section{Position of the particle}
\label{Sec:part_pos}

In order to derive the effective dynamics of the particle, we notice that the equation of motion \eqref{eq:field_four} of the Gaussian field is linear in $\phi$, enabling its formal integration. This yields
 \begin{equation}
     \phi_{\mathbf{q}}(t) = \int_{- \infty}^t \!\!\mathrm{d} t'\, e^{-\alpha_q(t - t')} \left[\lambda D_{\alpha}q^{\alpha} V_{\mathbf{q}} e^{-i \mathbf{q} \cdot \mathbf{X}(t')}  + \eta_{\mathbf{q}}(t')\right],
     \label{eq:field_by_X}
 \end{equation}
where we assumed that the initial conditions $\phi_{\mathbf{q}}(t_0)$ for the field configuration were imposed in the distant past, i.e., at $t_0 \to - \infty$,  deeming them inconsequential. 
In order to circumvent possible issues of equilibration of critical systems, we assume that the system is taken adiabatically to criticality after it has reached a stationary state, such that the limit $\xi \to \infty$ is taken after $t_0 \to -\infty$. 
%
%\ag{modified above a bit, check if you agree.}\dav{ok for me}
%
Notably, the field $\phi_{\mathbf{q}}(t)$ depends dynamically only on the position of the particle $\mathbf{X}(t)$ and the configuration of the noise $\eta_{\mathbf{q}}(t)$. Accordingly, the expression \eqref{eq:field_by_X} for the dynamics of the field can be substituted into Eq.~\eqref{eq:part}, describing the position of the particle, rendering an effective dynamics
 \begin{equation}    
     \dot{\mathbf{X}}(t) + \nu\kappa \left(\mathbf{X}(t) - \mathbf{v}t\right)\vartheta(-t) = \bm{F}\left[\{\mathbf{X}(t)\}\right] + \bm{\xi}(t),
     \label{eq:eff_dyn}
\end{equation}
  where 
\begin{equation}    
   \bm{F}\left[\{\mathbf{X}(t)\}\right] =  i \lambda\nu \int \frac{\mathrm{d}^d\mathbf{q}}{(2 \pi)^d} \mathbf{q}  V_{-\mathbf{q}} e^{i \mathbf{q} \cdot \mathbf{X}(t)} \left[ \phi_{\mathbf{q}}^{(0)}(t) + \lambda D q^{\alpha} V_{\mathbf{q}} \int_{-\infty}^t \mathrm{d}t' e^{- \alpha_q(t - t') - i\mathbf{q}\cdot\mathbf{X}(t')} \right],
   \label{eq:F}
\end{equation}    
and where
 \begin{equation}
     \phi_{\mathbf{q}}^{(0)}(t) = \int_{-\infty}^t \!\!\mathrm{d}t'\, e^{-\alpha_q (t - t')}\eta_{\mathbf{q}}(t')
     \label{eq:phi_0}
 \end{equation}
 is a colored Gaussian noise. 
 Note that the
 effective dynamics is not only non-Markovian, but also non-linear, and thus cannot be solved analytically. To overcome this difficulty, we will resort to the perturbative expansion introduced % an approximate method introduced \ag{why not simply "to the perturbative expansion introduced..."}\dav{I agree} 
 in Section~\ref{pert_exp}. 

\subsection{Perturbative expansion (weak-coupling approximation)}
\label{pert_exp}

In order to obtain analytical predictions for the effects of the field-particle interaction on the particle dynamics,
%tracer, 
following Refs.~\cite{Basu_2022,venturelli2023stochastic, Venturelli_2022, Venturelli_2022_2parts} we assume that $\lambda$ is a small parameter in terms of which one is allowed to perform a perturbative expansion:
\begin{gather}
    \mathbf{X}(t) = \mathbf{X}^{(0)}(t) + \lambda \mathbf{X}^{(1)}(t) + \lambda^2 \mathbf{X}^{(2)}(t) + \mathcal{O}(\lambda^3),
     \label{eq:particle_expansion} \\[1mm]
    \phi_{\mathbf{q}}(t) = \phi_{\mathbf{q}}^{(0)}(t) + \lambda \phi_{\mathbf{q}}^{(1)}(t)  + \lambda^2 \phi_{\mathbf{q}}^{(2)}(t) +\mathcal{O}(\lambda^3).
    \label{eq:field_expansion}
\end{gather}
In the studies cited above,
%In previous studies,
%
%\ag{Which ones? "In the studies cited above"?}
%
the expansion at the second order in $\lambda$ already proved sufficient to capture non-trivial effects on the particle motion. The results obtained within this weak-coupling approximation will be compared with those of numerical simulations of the system in, c.f., Fig.~\ref{fig:recoil}.

When the field and the particle are decoupled (i.e., for $\lambda=0$), the position of the latter is given by (see Eqs.~\eqref{eq:field_four} and~\eqref{eq:part})
\begin{equation}
    \mathbf{X}^{(0)}(t) =  \begin{cases}
        \mathbf{v}\left(t - \frac{1}{\nu \kappa}\right) + \int_{-\infty}^{t}\mathrm{d}t'\, e^{-\nu \kappa (t -t')}\bm{\xi}(t'), & \quad \mbox{for}\quad t\leq 0, \\[2mm]
        \mathbf{X}^{(0)}(0) + \int_{0}^{t}\mathrm{d}t'\, \bm{\xi}(t'), & \quad \mbox{for}\quad t >0, 
    \end{cases}
\end{equation}
where we chose the coordinate system such that at $t=0$ the minimum of the trap is located at the origin. 
In the expression above, for $t \leq 0$, one identifies the time integral of the noise as the Ornstein-Uhlenbeck process $\textbf{O}(t)$ with the relaxation time $\tau_X$ reported in Eq.~\eqref{eq:tau_xd}, and for $t>0$ as the Wiener process $\textbf{W}(t)$ with diffusion coefficient $\nu T$.
The symmetry of the particle dynamics reported below Eq.~\eqref{eq:force_def} implies that $\left\langle \mathbf{X}(t) \right\rangle$ is actually an even function of $\lambda$.  
Accordingly, the second-order correction to the decoupled dynamics is the lowest-order one with a non-vanishing mean, and it is given by
\begin{equation}
    \dot{\mathbf{X}}^{(2)}(t) + \vartheta(-t)\nu\kappa \mathbf{X}^{(2)}(t)  =  i \nu \int \frac{\mathrm{d}^d   \mathbf{q}}{(2\pi)^d} \mathbf{q}e^{i \mathbf{q}\cdot \mathbf{X}^{(0)}(t)}V_{-\mathbf{q}}\left[\phi_{\mathbf{q}}^{(1)}(t)  + i\mathbf{q}\cdot \mathbf{X}^{(1)}(t)  \phi_{\mathbf{q}}^{(0)}(t)\right].
    \label{eq:2nd_order_part}
\end{equation}
Here, we recognize the solution $\phi_{\mathbf{q}}^{(0)}(t)$ of the field dynamics with $\lambda=0$ as the colored noise introduced in Eq.~\eqref{eq:phi_0}.

\subsection{Steady-state average particle position}

Using the expression for $\langle e^{i\mathbf{q} \cdot (\mathbf{O}(t) - \mathbf{O}(t'))}\rangle$ reported in \cref{eq:app_OU_SF}, where $\mathbf{O}(t)$ is the 
%
%\ag{which ones? In short...} \textbf{[}
%
stationary Ornstein-Uhlenbeck process at time $t$ 
(see, e.g., Ref.~\cite{Basu_2022}), we calculate the lowest-order correction to the average position of the particle while it is subject to the harmonic confinement. The details of the computation are provided in  \ref{app:pert_cal}, while we report here only the final expression
\begin{equation}
    \left\langle \lambda^2 \mathbf{X}^{(2)} \right\rangle  =  \\ -\mathbf{\hat{e}}_1 \lambda^2  \frac{v}{\kappa} \int \frac{\mathrm{d}^d  \mathbf{q}}{(2 \pi)^d}\frac{q_1^2}{q^2 + r}|V_{\mathbf{q}}|^2\int_0^{\infty}\!\!\mathrm{d}s\, e^{- (\alpha_q - iq_1v)s - (T/\kappa)q^2(1 - e^{-\kappa \nu s})},
    \label{eq:shift}
\end{equation}    
where, for the sake of concreteness, we assumed $\mathbf{v} = \mathbf{\hat{e}}_1 v$, with $\mathbf{\hat{e}}_1$  being the unit vector in the direction of $x_1$. This time-independent average 
shift of the position of the particle from the center of the moving trap caused by the interaction with the field was derived previously in Ref.~\cite{venturelli2023stochastic}. 
Note that this coupling %to the field 
actually induces an additional friction, since the displacement in Eq.~\eqref{eq:shift} is negative, i.e., it occurs in the direction opposite to the velocity of the trap, as we prove in \ref{app:pos_shift}.
An analogous phenomenon was reported in Ref.~\cite{Demery_2019} while discussing a tracer driven in a colloidal bath. 
This friction can be understood as a result of the force  in Eq.~\eqref{eq:force_def} pulling the particle along the gradient of the steady-state field configuration 
forming behind it. A detailed description of this steady-state 
field profile is presented in  Section~\ref{sec:NESS} below.

In order to better understand the qualitative behavior of the average
particle displacement
in Eq.~\eqref{eq:shift}, we first consider the noiseless limit $T \to 0$ in which the fluctuations due to the thermal bath are negligible, and the right-hand side of the equation takes the simpler form
\begin{equation}
        \left\langle \lambda^2 \mathbf{X}^{(2)} \right\rangle_{T=0}  =  -\mathbf{\hat{e}}_1 \lambda^2  \frac{v}{\kappa} \int \frac{\mathrm{d}^d  \mathbf{q}}{(2 \pi)^d}D_{\alpha}q^{\alpha}|V_{\mathbf{q}}|^2 \frac{q_1^2 }{\alpha_q^2 + q_1^2 v^2}.
        \label{eq:shift_T0}
\end{equation}
Accordingly,  as a function of the velocity $v$, one observes first a linear growth of $\left\langle \lambda^2 \mathbf{X}^{(2)} \right\rangle_{T=0}$
followed by a decay to zero $\propto v^{-1}$ as $v$ is further increased\footnote{In fact, provided that the $\mathbf{q}$-integral in Eq.~\eqref{eq:shift_T0} converges, the right-hand side of this equation behaves like $v/(1 + Av^2)$, 
where $A$ is a constant --- for further discussion, see the supplemental material of Ref.~\cite{venturelli2023stochastic}.}. 
Accordingly, the shift vanishes both as $v \to 0$ and as $v \to \infty$. 
The former limit is easy to understand, because when the trap is not moving, the dynamics of the system is invariant under the inversion $x_1 \mapsto - x_1$, and thus the shift has to vanish. In the latter limit, the particle moves through the medium too quickly for the field to react to its presence, and thus the dynamics of the field effectively decouples from that of the particle. A detailed analysis of the case of dissipative (model A) dynamics in spatial dimension $d=1$ is presented in  the supplemental material of Ref.~\cite{venturelli2023stochastic}.

%
% QUIQUIQUIQUIQUIQUIQUI
%

\subsection{Recoil}
\label{sec:recoil}

\begin{figure}
    \centering
        \includegraphics[width=0.5\linewidth]{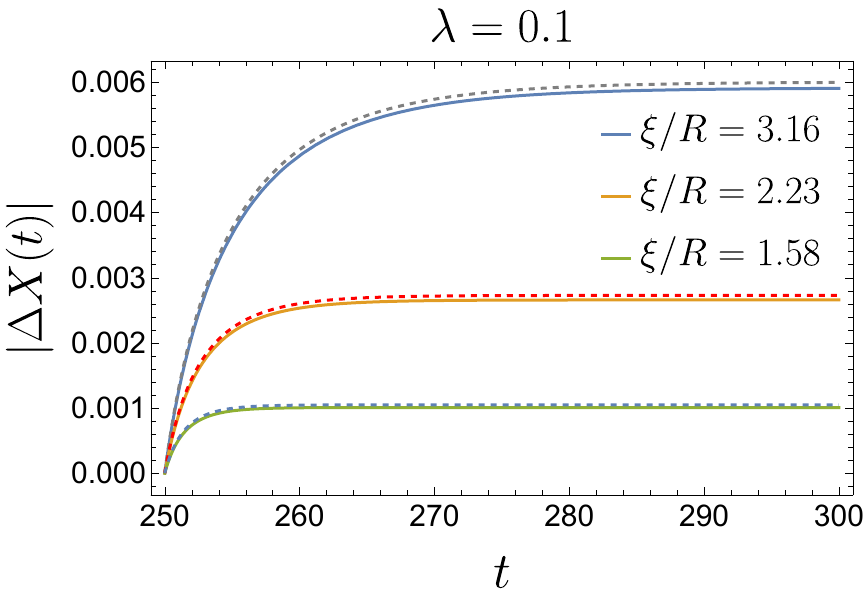}
\caption{Dependence of the recoil amplitude $|\Delta X|$ on time,  for the non-critical model~A  in spatial dimension $d=1$, in the absence of noise ($T=0$) and for various values of the correlation length $\xi$ of the field. The interaction of the particle with the field is assumed to be Gaussian as in Eq.~\eqref{eq:Gauss_kernel} with $R=1$, while the other parameters of the model are $\kappa = 0.5$, $D=1$, $\nu =1$, $\lambda=0.1$, and $v=1$. The solid lines correspond to the predictions of the perturbative calculation reported in~\cref{zeroT_rec}, while the dashed lines are the result of a numerical simulation in which the particle was dragged through the medium for $t \in [0, 250]$, before being released from its confinement at time $t = 250$.
}
\label{fig:recoil}
\end{figure}

The main focus of this work is on the motion of the particle after it has been  released from the harmonic trap. It turns out that the particle exhibits recoil, i.e., it moves in the direction opposite to the drag, in a manner similar to what is observed experimentally in viscoelastic media~\cite{Gomez-Solano_2015, Robertson-Anderson_2018,Ginot_2022_rec, Cao_2023}. 
We note that this backwards motion arises as a result of the field-particle interaction, and vanishes for $\lambda=0$: indeed, due to the symmetry of the particle's dynamics reported under \cref{eq:force_def}, one can verify that
%
%Here we denote the time-dependent recoil by $\mathbf{ \Delta X}^{(2)}(t) = \mathbf{X}^{(2)}(t) - \mathbf{X}^{(2)}(0)$
\begin{equation}
    \left \langle\mathbf{X}(t) \rangle - \langle\mathbf{X}(0) \right \rangle = \left \langle\lambda^2\mathbf{ \Delta X}^{(2)}(t) \right \rangle + {\cal O}(\lambda^4),
\end{equation}
where we denoted the time-dependent recoil by $\lambda^2\mathbf{ \Delta X}^{(2)}(t)$.
%
%\ag{Perhaps here we should also state more clearly that $\mathbf{X}(t) - \mathbf{X}(0) = \lambda^2\mathbf{ \Delta X}^{(2)}(t) + {\cal O}(\lambda^4)$, i.e., that the recoli at order $\lambda^0$ vanishes, right? }
%
We postpone the derivation of the recoil to \ref{app:pert_cal} and present here the final expression for its average:
\begin{multline}    
    \left\langle\lambda^2 \mathbf{ \Delta X}^{(2)}(t) \right\rangle  =  \\ -\mathbf{\hat{e}}_1 \lambda^2  \nu v \int \frac{\mathrm{d}^d  \mathbf{q}}{(2 \pi)^d}\frac{q_1^2}{q^2 + r}|V_{\mathbf{q}}|^2\frac{1 - e^{-(\alpha_q + q^2 \nu T)t}}{\alpha_q + q^2 \nu T}\int_0^{\infty}\dd{s} e^{- (\alpha_q - iq_1v)s - (T/\kappa)q^2(1 - e^{-\kappa \nu s})}.
    \label{eq:recoil_full}
\end{multline}
Note that Eq.~\eqref{eq:recoil_full}, as a function of time $t$, possesses the mathematical structure of a weighted exponential decay towards its final value. However, when the relaxation time $\tau_\phi(q)$  of the small-wavevector modes $\phi_{\mathbf{q} \simeq \mathbf 0}$ diverges (see Eq.~\eqref{eq:alpha_q}), i.e., generically for model B and for model A at criticality, algebraic decays are expected. 
Note also that a non-vanishing temperature $T$ causes a renormalization of the (inverse) relaxation times $\alpha_q$ according to 
  \begin{equation}
      \alpha_q \mapsto \alpha_q' = \alpha_q + q^2 \nu T ,
      \label{eq:alphaq_prim}
  \end{equation}
which is equivalent to  $D_0 \mapsto D_0 + \nu T $ for model A, and  $r \mapsto r + \nu T/D_2$ for model B. Consequently, a finite temperature $T$ suppresses critical fluctuations in the case of model B, as it effectively drives the system away from the critical point $r=0$.
The short-time behavior of the recoil can be easily obtained from Eq.~\eqref{eq:recoil_full} by approximating $1 - e^{-\alpha'_t t} \simeq  \alpha'_q t$, which renders  
\begin{equation}
\left\langle\lambda^2 \mathbf{ \Delta X}^{(2)}(t) \right\rangle   \simeq  -\mathbf{\hat{e}}_1 \lambda^2  
\nu v  t\int \frac{\mathrm{d}^d  \mathbf{q}}{(2 \pi)^d}\frac{q_1^2}{q^2 + r}|V_{\mathbf{q}}|^2\int_0^{\infty}\!\!\mathrm{d}s\, e^{- (\alpha_q - iq_1v)s - (T/\kappa)q^2(1 - e^{-\kappa \nu s})}.
\label{eq:rec_smallt}
\end{equation} 
In passing, we note that this expression coincides with Eq.~\eqref{eq:shift} for the steady-state shift  $\left\langle \lambda^2 \mathbf{X}^{(2)}(t) \right\rangle$ multiplied by the dimensionless time $t/\tau_X$, where $\tau_X$ is the relaxation time of the particle in the trap, introduced in Eq.~\eqref{eq:tau_xd}. 
This is because, before the particle is released, the field-induced friction experienced by the particle is counteracted by the force $\mathbf{f}_0 = \kappa  \langle \lambda^2 \mathbf{X}^{(2)}(t<0)\rangle$  generated by the trap as a consequence of the additional shift $\langle \lambda^2 \mathbf{X}^{(2)}(t<0)\rangle$ from its center, maintaining mechanical equilibrium in the co-moving reference frame in which the trap is at rest. 
When the particle is released from confinement, the force $\mathbf{f}_0$ is exerted on the particle by the field in its initial configuration at $t=0$. % due to its slow relaxation. 
This force, due to the overdamped nature of the particle dynamics characterized by the mobility $\nu$, gives an initial velocity $\mathbf{v}_0 = \nu \mathbf{f}_0$ to the particle, causing a recoil which initially grows as $\mathbf{v}_0 t = \nu \kappa t \langle \lambda^2 \mathbf{X}^{(2)}(t<0)\rangle = (t/\tau_X)\langle \lambda^2 \mathbf{X}^{(2)}(t<0)\rangle $, see Eq.~\eqref{eq:tau_xd}, as noted above. 
In the noiseless limit $T\to 0$, the dynamics is deterministic and Eq.~\eqref{eq:recoil_full} becomes
\begin{equation}\left\langle\lambda^2 \mathbf{ \Delta X}^{(2)}(t) \right\rangle   =   -\mathbf{\hat{e}}_1 \lambda^2  \nu v \int \frac{\mathrm{d}^d  \mathbf{q}}{(2 \pi)^d}\frac{q_1^2}{q^2 + r}|V_{\mathbf{q}}|^2\frac{1 - e^{-\alpha_qt}}{\alpha_q ^2 + q_1^2 v^2},
\label{zeroT_rec}
\end{equation}
which does not depend on the trap strength $\kappa$
(which controls the spatial extent of the thermal fluctuations of the particle position while it is driven by the trap). 
The corresponding recoil is plotted as a function of time in Fig.~\ref{fig:recoil} 
for the non-critical model~A in $d=1$ and with the Gaussian interaction kernel given in Eq.~\eqref{eq:Gauss_kernel}, along with the results of a numerical simulation of the 
system.
For the data presented, the relative difference between the analytic prediction and the results of numerical simulations is of the order of 4$\%$. This difference increases upon increasing the coupling strength $\lambda$, reaching $\simeq 6\%$ for $\lambda = 0.5$.

%%%%%%%%%%%%%%%%%%%%%%%%%%%%%%%%%%%%%%%%%
%%%%%%%%%%%%%%%%%%%%%%%%%%%%%%%%%%%%%%%%%
%%%%%%%%%%%%%%%%%%%%%%%%%%%%%%%%%%%%%%%%%
%%%%%%%%%%%%%%%%%%%%%%%%%%%%%%%%%%%%%%%%%

It is convenient to express the recoil in the noiseless limit, see Eq.~\eqref{zeroT_rec}, in terms of a scaling function $\mathcal{F}^{A/B}$, i.e.,
\begin{equation}
    \left\langle\lambda^2 \mathbf{ \Delta X}^{(2)}(t) \right\rangle = -\mathbf{\hat{e}}_1 g\frac{v \tau_R^2}{\tau_X}\mathcal{F}^{A/B}\left( \frac{R}{\xi}, 
    \frac{t}{\tau_R}, \frac{v \tau_R}{R}\right),
    \label{zeroT_rec_scale_def}
\end{equation}
where 
\begin{equation}
\mathcal{F}^{A/B}\left(1/\tilde{\xi}, 
\tilde t, \tilde v\right) =  \int\frac{\mathrm{d}^d  \mathbf{q}}{(2 \pi)^d}\frac{q_1^2}{q^2 + 1/\tilde{\xi}^2}|\hat{V}_{\mathbf{q}}|^2\frac{1 - e^{-\tilde{t} q^{\alpha}\left(q^2 +1/\tilde{\xi}^2\right)}}{q^{2\alpha}\left(q^2 +1/\tilde{\xi}^2\right)^2 + \tilde{v}^2q_1^2},
\label{zeroT_rec_scale}
\end{equation}
and where $\hat{V}_{\mathbf{q}}$ is the Fourier transform (see \cref{eq:FT}) of the interaction kernel rescaled by the particle's effective radius introduced after \cref{eq:Hint}. 
The superscript $A$ or $B$ to $\mathcal{F}$ and to similar scaling functions introduced further below indicate that they refer to the case of model A ($\alpha=0$) or model B ($\alpha = 2$).  
The prefactor $g v \tau_R^2/\tau_X$ in \cref{zeroT_rec_scale_def}
has the physical dimensions of a length, and it is expressed in terms of a dimensionless coupling constant 
\begin{equation} 
g = \frac{\lambda^2}{\kappa R^ d}, 
\label{eq:def-g}
\end{equation}
of the timescale $\tau_X$  defined in Eq.~\eqref{eq:tau_xd}, and of
\begin{equation}
    \tau_{R} = 
    %\frac{R^{z}}{D_{\alpha}},
    R^{z}/D_{\alpha},
\label{eq:def-tau-R}
\end{equation}
which  characterizes the relaxation time of the critical field over a distance of the order of the particle size $R$ (its definition follows from Eq.~\eqref{eq:alpha_q} with $r = 0$ and $q \sim 1/R$). 
In the convention introduced in Eq.~\eqref{zeroT_rec_scale_def}, distances are measured in terms of 
$R$, and times in terms of $\tau_R$, allowing for the introduction of the dimensionless correlation length and the dimensionless time according to 
\begin{equation}
    \tilde{\xi} = \xi/R \quad \mbox{and} \quad \tilde{t} = t/\tau_R.
    \label{eq:dimless_var}
\end{equation}
We note that the last argument of the scaling function in Eq.~\eqref{zeroT_rec_scale_def}, i.e., the dimensionless velocity
    \begin{equation}
             \tilde{v} \equiv v\tau_{R}/R =2\mathrm{Wi} 
     =  \left(R/l_{v, \alpha}\right)^{\alpha + 1},
        \label{eq:Weissenberg}
    \end{equation}
can be expressed in terms of
the Weissenberg number $\mathrm{Wi}$ 
used in the context of micro-rheological experiments in viscoelastic media~\cite{Berner_2018, Poole_2012},
which compares the typical times describing the medium relaxation and its deformation. Indeed, the dragged particle deforms the medium over the length of its diameter in a time $2R/v$. 
Note that the length scale % \dav{Sorry if I realized it just now, but why did we choose this notation? It bears no connection to $\xi$, so why not $l_{v, \alpha}$ or something less confusing?}
\begin{equation}
l_{v, \alpha} = \left( 
%\frac{D_{\alpha}}{v}
D_{\alpha}/v
\right)^{1/(\alpha + 1)}
\label{eq:xi-v}
\end{equation}
arises naturally while describing the steady-state field configuration at criticality, as discussed in, c.f.,
\cref{sec:ss}.
%Eq.~(\ref{eq:xi_v_def}). \dav{broken link}

We note that in the limit of instantaneous equilibration of the field, corresponding \rev{(for non-critical model A) to $D_{0} \to \infty$} 
%to $D_{\alpha} \to \infty$ 
or, equivalently, $\tau_R \to 0 $,  the expression for the recoil in Eqs.~\eqref{zeroT_rec_scale_def} and~\eqref{zeroT_rec_scale} generically vanishes;  in this \textit{adiabatic} limit, 
the timescale separation is recovered\footnote{\rev{More precisely, timescale separation requires that the relaxation time of the slowest mode of the field (i.e., the one corresponding to the smallest wavevector, see Eq.~\eqref{eq:alpha_q}) vanishes. Hence, in the case of model B and critical model A, the presence of zero modes with diverging relaxation time precludes a \textit{bona-fide} adiabatic limit (see
Refs.~\cite{Venturelli_2022,Venturelli_Gross_2022} for further discussions).}}.
We therefore emphasize that the presence of the recoil stems from the fact that the dynamics of the particle and of the field occur over comparable timescales.

The convention adopted here allows us to investigate the following relevant limits of Eq.~\eqref{zeroT_rec_scale_def}:  (i) the limit of small velocities $v$; (ii) the long-time limit which we denote by
\begin{equation}
\mathcal{R}^{A/B}\left(1/\tilde{\xi
} , \tilde{v}\right) = \lim_{\tilde{t} \to \infty} \mathcal{F}^{A/B}\left( 1/\tilde{\xi}, 
\tilde{t}, \tilde{v}\right),
\label{eq:range_def}
\end{equation}
where we will henceforth refer to the quantity $\tilde{v} \times \mathcal{R}^{A/B}(1/\tilde{\xi}, \tilde{v})$ as the \textit{dimensionless recoil range} because it describes the final particle position (recall that $\left\langle \lambda^2\Delta \mathbf{X}^{(2)} \right\rangle \propto v\mathcal{F}^{A/B}$ in~\cref{zeroT_rec_scale_def});
and (iii) the critical limit $\xi \to \infty$.
Concerning (i), in the case of finite $\xi$ and $t$, at small velocities, the recoil is characterized by a linear growth upon increasing $v$, as indicated by the prefactor on the right-hand side of Eq.~\eqref{zeroT_rec_scale_def}. 
However, when at least one of the relevant tunable scales in Eq.~\eqref{zeroT_rec_scale_def} diverges (i.e., either $t$ or $\xi$), the scaling function $\mathcal{F}^{A/B}$ acquires a non-trivial behavior at small $v$, which is analyzed by means of the Mellin transform in \ref{app:small_v} and is reported in
Tab.~\ref{tab:small_v_scale}.   

%%%%%%%%%%%%%%%%%%%%%%%%%%%%%%%%%%%%%%%%%%%%%%
\begin{table}
    \centering
\begin{center}
\vspace{.2cm}
\begin{tabular}{ | l | l | ll | } 
   \cline{3-4}
   \multicolumn{2}{c|}{} & &  \\[-5mm] 
   \multicolumn{2}{c|}{ {$\mathcal{F}^{A/B}(1/\tilde{\xi},\tilde{t},\tilde{v}\to 0)$}}& $\tilde{\xi} < \infty$  & $\tilde{\xi} \to \infty$   \\ 
  \hline
  \multirow{2}{*}{$\alpha=0$} &$ \tilde{t} < + \infty$ & $\sim 1$ & $\sim 1 +  f_{0}(d; \tilde{t})\tilde{v}^{d-2}$ \\ 
  
  &$ \tilde{t} \to \infty$ & $\sim 1$ & $\sim 1 + 
   g_{0}(d)\tilde{v}^{d-4}$ \\ 
  \hline
  \multirow{2}{*}{$\alpha=2$} &$ \tilde{t} < + \infty$ & $\sim 1$ & $\sim 1 +  f_{2}(d; \tilde{t})\tilde{v}^{(d-4)/3}$ \\ 
  &$ \tilde{t} \to \infty$ & $\sim 1 + 
  h(d; \tilde{\xi
  })\tilde{v}^{d-2}$ & $\sim 1 + 
  g_{2}(d)\tilde{v}^{(d-8)/3}$ \\ 
  \hline
\end{tabular}
\end{center}
    \caption{
        %\ag{Edited: check if you agree.} 
    Behavior of the scaling function $\mathcal{F}^{A/B}(1/\tilde{\xi},\tilde{t},\tilde{v})$ defined in Eq.~\eqref{zeroT_rec_scale_def} for $\tilde{v}\to 0$. The prefactors $f_{0,2}(d; \tilde{t})$, $g_{0,2}(d)$, and $h(d; \tilde{\xi})$, which depend on the dimensionless scaling variables $\tilde{t}$ and $\tilde{v}$ defined in \cref{eq:dimless_var,eq:Weissenberg},
    are determined by means of the Mellin transform in \cref{eq:app_mellin_t_fin} and \cref{eq:app_mellin_t_inf}. Physically, the generic non-analytic dependence of the scaling function on $\tilde{v}$ is due to the fact that the relaxation time of the field, whose inverse is given by Eq.~\eqref{eq:alpha_q}, diverges. The dependence on $d$ of the power of $\tilde{v}$ appearing in the table originates from the infrared divergence of the $\mathbf{q}$-integral in \cref{zeroT_rec_scale}, which is regularized for sufficiently large $d$. 
    }
    \label{tab:small_v_scale}
\end{table} 
%%%%%%%%%%%%%%%%%%%%%%%%%%%%%%%%%%%%%%%%%%%%%%

Depending on the type of dynamics and on the space dimensionality $d$, this might even result in a diverging recoil range as $v \to 0$: this occurs for 
critical model A in $d < 3$ and for critical model B in %$d < 1$ and 
$d <5$ 
(recall the presence of $v$ in the prefactor in Eq.~\eqref{zeroT_rec_scale_def}). 
For critical model A in $d=3$ and model B in $d=1$ and $d=5$ with $r>0$ and $r=0$, respectively, the limit $v \to 0$ yields a finite recoil range. 
 The physical interpretation of this result is that, in a system characterized by diverging relaxation time, any small deformation rate 
 %characterized 
 induced
 by the driving velocity of the particle is still much larger than the relaxation rate of the medium. Hence, for any infinitesimal, yet nonzero $\tilde{v}$, the recoil range may attain a finite value or even diverge.
 A similar behavior for $\left\langle \lambda^2 X^{(2)} (t<0)\right\rangle$ given by Eq.~\eqref{eq:shift} was reported in the supplemental material of Ref.~\cite{venturelli2023stochastic}.

In order to discuss in more detail the
asymptotic long-time
behavior of the recoil,
which is the object of the next section, 
the kernel $V(\mathbf{x})$ describing the field-particle interaction is henceforth assumed to have the Gaussian form reported in Eq.~\eqref{eq:Gauss_kernel}, unless differently stated.

\subsection{Long-time behavior}
\label{sec:asympt}
In this section we analyze the behavior of the scaling function of the recoil defined in \cref{zeroT_rec_scale}, starting with the long-time asymptotics of the critical recoil (i.e., the average particle recoil when $\tilde \xi \to \infty$), and following with the description of $\mathcal{R}^{A/B}$ defined in \cref{eq:range_def} for a large but finite correlation length. The corresponding calculations are outlined in \ref{app_lin_scale}. Moreover, explicit expressions for the critical recoil in the case of model~A dynamics with $d \in \{1, 2, 3\}$ are reported in \ref{app:rec_full}. Here we summarize the key results of our analysis.

For $d >2$, at long times the critical recoil
satisfies
\begin{equation}
    \mathcal{F}^{A/B}\left( 0, \tilde{t}, \tilde{v}\right) \sim   \mathcal{R}^{A/B}\left(0, \tilde v\right) + \frac{\tilde{t}^{(2-d)/z}}{\tilde{v}^2} |\hat{V}_0|^2   \mathcal{C}_{\alpha}(d), \quad \mbox{for} \quad  \tilde{t} \gg 1, 
    \label{eq:crit_recoil_asympt}
\end{equation}
where the dynamical critical exponent $z$ of the field was introduced after Eq.~\eqref{eq:field_rec}, 
and a finite dimensionless velocity $\tilde v$ is assumed.
Here, $\mathcal{C}_{\alpha}(d)$ is a constant determined solely by the spatial dimensionality and by the dynamics of the field, 
whose expression is
reported in Eq.~\eqref{app:c_alpha}. Interestingly, $\mathcal{R}^{A/B}(0, \tilde{v})$ does not exist for $d\leq 2$ due to infrared divergences: physically, this corresponds to the particle moving backwards indefinitely due to diverging relaxation times. The presence of such a diverging recoil range for $d \leq 2$ can also be understood as a consequence of a peculiarity of the scalar Gaussian field theory at the critical point, discussed in more detail in \cref{sec:ss} (see, c.f., \cref{eq:scalarFT}).  
Notably, for $d>2$, $\mathcal{C}_{\alpha}(d)$ is negative and the recoil converges to its final value 
$\mathcal R^{A/B}$ 
from below, with a power-law dependence on time with exponent $(2 - d)/z$.  
However (as shown in \ref{app:rec_full}, where the full expressions for the scaling function in the case of critical dissipative dynamics with $d \in \{1,2,3 \}$ are provided), in $d=1$, the long-time asymptotic behavior of the recoil is governed by the second term on the right-hand side of Eq.~\eqref{eq:crit_recoil_asympt}, i.e., it grows linearly as a function of time with the corresponding prefactor 
$\propto \mathcal{C}_0(1)$ which is positive. 
In $d=2$, instead, both $\mathcal{C}_{0}(d)$ and $\mathcal{C}_{2}(d)$ have simple poles as functions of $d$, and thus the recoil increases logarithmically upon increasing time. 
We emphasize that the second term on the right-hand side of Eq.~(\ref{eq:crit_recoil_asympt}) is largely universal, as it depends only on the behavior of the interaction kernel $\hat{V}_{\mathbf{q}}$ at 
wavevector $\mathbf{q}=\bm 0$, 
rendering it insensitive to the details of the field-particle interaction (as long as it is linear, translationally and rotationally invariant).

\begin{table}
    \centering
\begin{center}
\vspace{.2cm}
\begin{tabular}{ | l | c | c| c | } 
   \cline{3-4}
   \multicolumn{2}{c|}{} & &  \\[-5mm] 
   \multicolumn{2}{c|}{ {$\mathcal{F}^{A/B}\left(1/\tilde{\xi},\tilde{t} \gg 1,\tilde{v}\right)$}}& $\tilde{\xi} < \infty$  & $\tilde{\xi} \to \infty$   \\ 
  \hline
  \multirow{4}{*}{$\alpha=0$} &\multirow{2}{*}{$ d \leq 2$} & \multirow{4}{*}{$\sim \mathcal{R}^A\left(1/\tilde{\xi},\tilde{v}\right)-  \frac{|\hat{V}_0|^2}{2(4 \pi)^{d/2} }  \frac{\tilde{\xi}^{6} e^{-\tilde{t}/\tilde\xi^2}}{ \tilde{t}^{d/2+1}}$} & \multirow{2}{*}{ $ \sim\frac{\tilde{t}^{(2-d)/2}|\hat{V}_0|^2}{\tilde{v}^{2}}    \mathcal{C}_{0}(d)$} \\ 
  &  &  &   \\ & \multirow{2}{*}{$ d > 2$} & & \multirow{2}{*}{$\sim\mathcal{R}^A\left(0,\tilde{v}\right) + \frac{|\hat{V}_0|^2}{\tilde{t}^{(d-2)/2}\tilde{v}^{2}}    \mathcal{C}_{0}(d)$}
  \\ &  & &
  \\ \hline \multirow{4}{*}{$\alpha=2$} & \multirow{2}{*}{$ d \leq 2$} & \multirow{4}{*}{$\sim\mathcal{R}^{B}\left( 1/\tilde{\xi},  \tilde{v}\right)
-  \frac{|\hat{V}_0|^2}{(4 \pi)^{d/2} }  \frac{\tilde{\xi}^{d + 2}}{\tilde{v}^2 \tilde{t}^{d/2}}$} & \multirow{2}{*}{ $ \sim\frac{\tilde{t}^{(2-d)/4}|\hat{V}_0|^2}{\tilde{v}^{2}}    \mathcal{C}_{2}(d)$}  \\ 
   & & &   \\ 
  &$ \multirow{2}{*}{$ d > 2$}$ & &  \multirow{2}{*}{$\sim \mathcal{R}^B\left(0,\tilde{v}\right) + \frac{|\hat{V}_0|^2}{\tilde{t}^{(d-2)/4}\tilde{v}^{2}}    \mathcal{C}_{2}(d)$} \\ 
  & & & \\ 
  \hline
\end{tabular}
\end{center}
    \caption{Long-time behavior of the scaling function $\mathcal{F}^{A/B}\left(1/\tilde{\xi},\tilde{t},\tilde{v}\right)$ in \cref{zeroT_rec_scale}.  The divergence  of the recoil range for $\tilde t \to \infty$, occurring if $\tilde{\xi} \to \infty$ and $d \leq 2$, indicates an endless backward motion of the particle after its release from the trap (see the discussion in \cref{sec:ss,sec:asympt}). Note that $\mathcal{C}_{0,2}(d)$ is positive for $d < 2$, and negative for $d>2$.}
    \label{tab:range}
\end{table}

\begin{figure}
    \centering
    \begin{tabular}{cc}
        \includegraphics[width=0.47\linewidth]{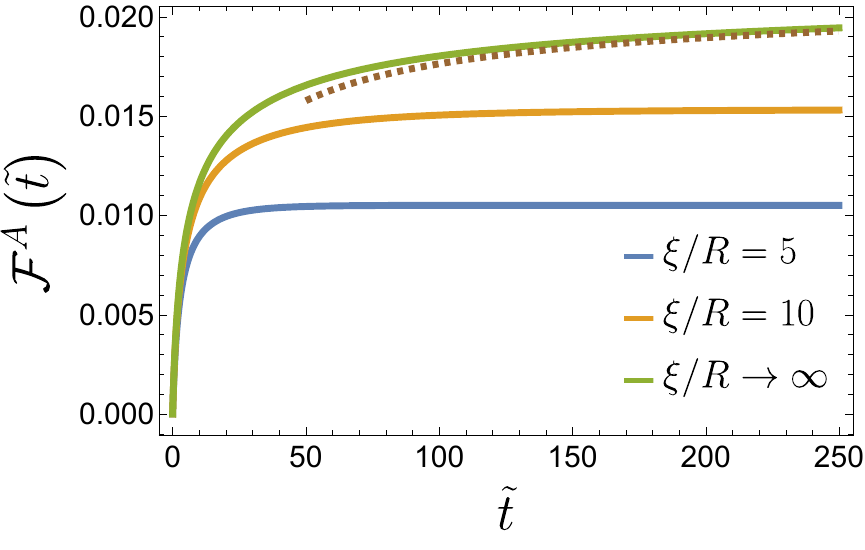} &
        \includegraphics[width=0.47\linewidth]{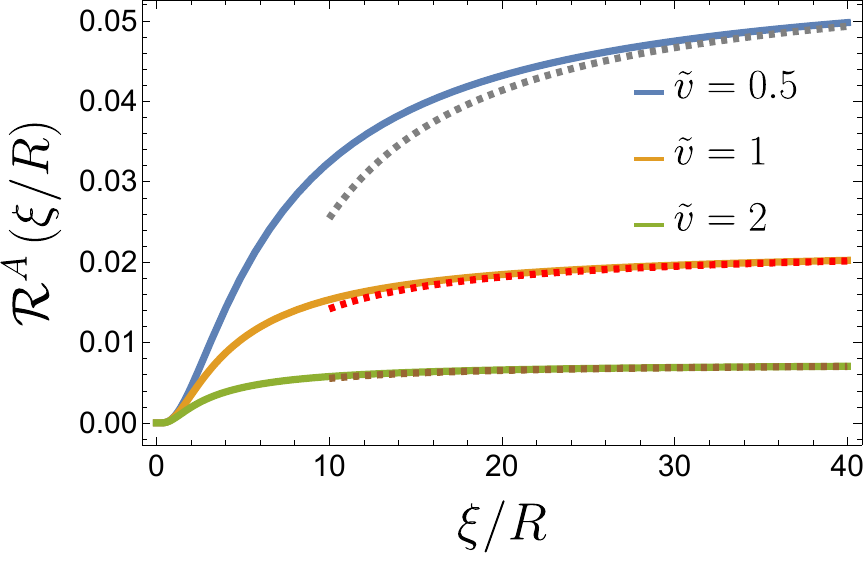}
        % \\[-2mm]
        % (a) & (b)
    \end{tabular}
        \put(-420,-55){(a)}
    \put(-205,-55){(b)}
\caption{Scaling function $\mathcal F^A$ of the recoil for model A in spatial dimension $d=3$, see \cref{zeroT_rec_scale}.
Panel~(a) shows $\mathcal F^A$
as a function of the dimensionless time $\tilde t$ with $\tilde v=1$, and various values of the dimensionless correlation length $\xi/R$. The dashed line corresponds to the long-time behavior at the critical point reported  in \cref{eq:crit_recoil_asympt}, where $z = 2$. 
Panel~(b) shows the dependence of the long-time limit $\mathcal R^A$  of the scaling function $\mathcal F^A$  of the recoil (see \cref{eq:range_def}) as a function of $\xi/R$, for various values of the scaled driving velocity $\tilde{v}$. The behaviors of these curves for large values of $\xi/R$, given by \cref{eq:range_asympt}, are reported as dashed lines.
}
\label{fig:recoil-range}
\end{figure}

\begin{figure}
    \centering
    \begin{tabular}{cc}
        \includegraphics[width=0.47\linewidth]{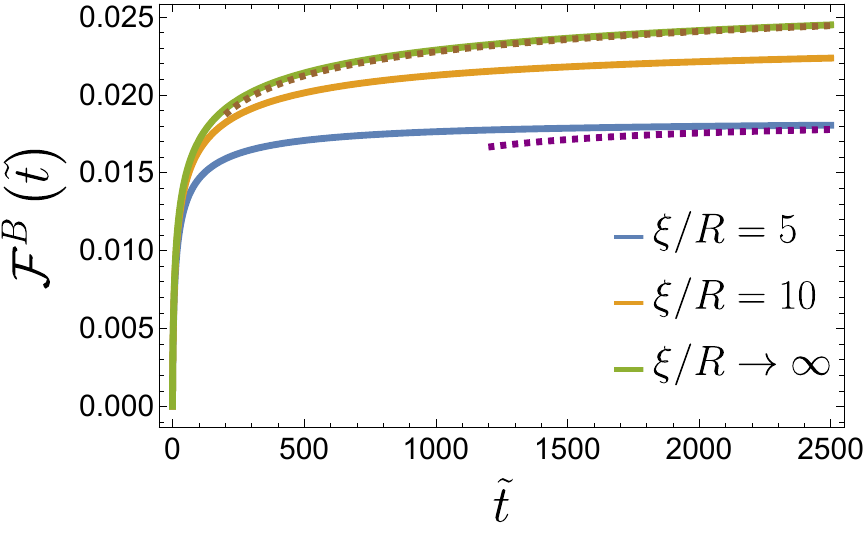} &
        \includegraphics[width=0.46\linewidth]{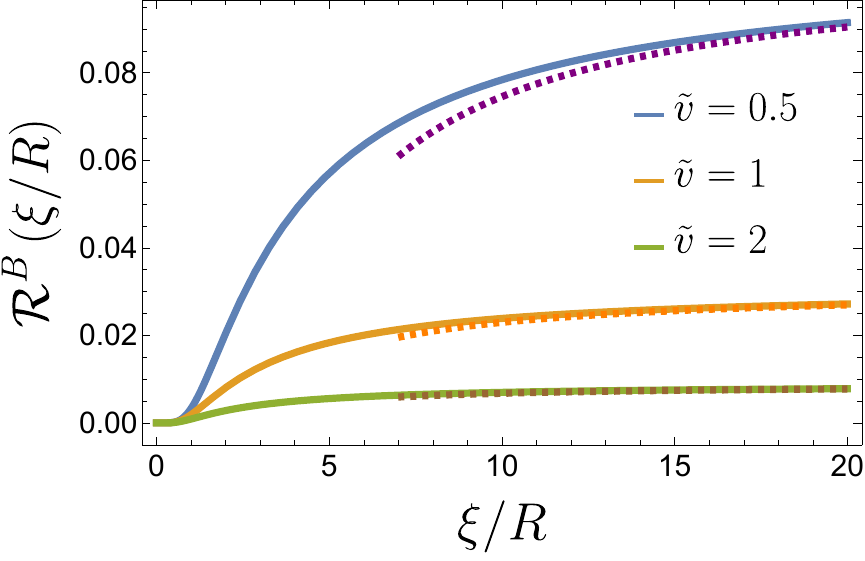}
        %\\[-2mm]
        %(a)&(b)
    \end{tabular}
    \put(-420,-55){(a)}
    \put(-205,-55){(b)}
\caption{Scaling function $\mathcal F^B$ of the recoil for model B in spatial dimension $d=3$, see \cref{zeroT_rec_scale}.
Panel (a) shows $\mathcal F^B$
as a function of the dimensionless time $\tilde t$ with $\tilde v=1$, and various values of the dimensionless correlation length $\xi/R$. The dashed brown line corresponds to the long-time behavior at the critical point reported  in \cref{eq:crit_recoil_asympt}, where $z = 4$, and the dashed purple line corresponds to the long-time behavior in the non-critical case reported in \cref{eq:subcrit_B_recoil_asympt}.
Panel (b) shows the dependence of the long-time limit $\mathcal R^B$  of the scaling function $\mathcal F^B$  of the recoil (see \cref{eq:range_def}) as a function of $\xi/R$, for various values of the scaled driving velocity $\tilde{v}$. The behaviors of these curves for large values of $\xi/R$, given by \cref{eq:range_asympt}, are reported as dashed lines.
}
\label{fig:recoil-rangeB}
\end{figure}

Let us now turn to non-critical systems, where $\mathcal{R}^{A/B}$ turns out to be always finite. In particular,
in the case of dissipative dynamics, the particle relaxes exponentially to its final position according to 
\begin{equation}
    \mathcal{F}^{A}\left( 1/\tilde{\xi}, \tilde{t}, \tilde{v}\right) \sim  \mathcal{R}^{A}\left( 1/\tilde{\xi},  \tilde{v}\right)
-  \frac{|\hat{V}_0|^2}{2(4 \pi)^{d/2} }  \frac{\tilde{\xi}^{6} e^{-\tilde{t}/\tilde\xi^2}}{ \tilde{t}^{d/2+1}}  , \qquad \mbox{for}\qquad \tilde{t} \gg 1.
\label{eq:subcrit_A_recoil_asympt}
\end{equation}
In the case of model B, instead, the particle recoil attains its maximal value following an algebraic approach according to
\begin{equation}
    \mathcal{F}^{B}\left( 1/\tilde{\xi}, \tilde{t}, \tilde{v}\right) \sim  \mathcal{R}^{B}\left( 1/\tilde{\xi},  \tilde{v}\right)
-  \frac{|\hat{V}_0|^2}{(4 \pi)^{d/2} }  \frac{\tilde{\xi}^{d + 2}}{\tilde{v}^2 \tilde{t}^{d/2}}  , \qquad \mbox{for} \qquad \tilde{t} \gg 1.
\label{eq:subcrit_B_recoil_asympt}
\end{equation}
A summary of the long-time behavior of the scaling function $\mathcal{F}^{A/B}$ is presented  Tab.~\ref{tab:range}. We point out that when the dynamics is characterized by diverging relaxation times, and thus the approach to the maximal recoil occurs algebraically in time,
the difference $\mathcal{F}^{A/B} - \mathcal{R}^{A/B}$ between the corresponding scaling functions scales with the dimensionless velocity  $\tilde{v}$ like $\sim \tilde{v}^{-2}$, see Eqs.~\eqref{eq:crit_recoil_asympt} and \eqref{eq:subcrit_B_recoil_asympt}. 
In the case of non-critical model A (which is characterized by a finite relaxation time), instead, the corresponding difference in Eq.~\eqref{eq:subcrit_A_recoil_asympt} does not depend on the velocity.
These two dependences on the velocity coincide with  the behavior of $\mathcal{F}^{A/B}$ in the  regime of $\tilde{v}\gg 1$ (far from equilibrium) and $\tilde{v}\ll 1$ (close to equilibrium), respectively.
 $\mathcal F^{A/B}$ 
for $d=3$ as a function of time are plotted in 
Figs.~\ref{fig:recoil-range}(a) and \ref{fig:recoil-rangeB}(a) for model A and B, respectively.
The asymptotic curves introduced in Eqs.~\eqref{eq:crit_recoil_asympt} and  \eqref{eq:subcrit_B_recoil_asympt} are reported as dashed lines.

%%%%%%%%%%%%%%%%%%%%%%%%%%%%%%%%%%%%%%%%%%%%

%%%%%%%%%%%%%%%%%%%%%%%%%%%%%%%%%%%%%%%%%%%%

For $d < 2$, it is instructive to investigate the asymptotic behavior of $\mathcal{R}^{A/B}$ for large but finite values of the correlation length $\xi$. It turns out that the expressions corresponding to model A and B actually coincide, 
and that they are given by
\begin{equation}
\mathcal{R}^{A/B}\left( 1/\tilde{\xi}, \tilde{v} \right) \sim |\hat{V}_0|^2\frac{\Gamma(1 -d/2) }{(4 \pi)^{d/2}}  \frac{\tilde{\xi}^{2-d}}{\tilde{v}^2}, \qquad \mbox{for}\qquad \tilde{\xi}
\gg 1.
\label{eq:range_asympt}
\end{equation}
Recall that the definition of the dimensionless velocity $\tilde{v}$ in  \cref{eq:Weissenberg} differs for models A and B. %
Note that the exponent that describes the power-law divergence of the recoil upon increasing the correlation length $\xi$ in Eq.~\eqref{eq:range_asympt} agrees with that reported in Eq.~\eqref{eq:crit_recoil_asympt}, which describes the long-time asymptotics of the critical recoil, taking into account the scaling relation $t \sim \xi^z$, where $z$ is the dynamical critical exponent. 
It turns out that, for $d = 3$, Eq.~\eqref{eq:range_asympt} describes the asymptotic $\sim 1/\xi$  convergence of $\mathcal{R}^{A/B}$ towards its maximal value obtained in the limit of $\xi \to \infty$, which we show by plotting it as dashed curves in Figs.~\ref{fig:recoil-range}(b) and~\ref{fig:recoil-rangeB}(b). %  \textbf{[!!!]}

In passing, we note that while describing the final position of the particle after the recoil, the limits $v \to 0 $ and $\xi \to \infty$ do not commute, see \cref{eq:range_asympt}. In fact, it turns out that the final particle position depends on the driving velocity via the variable
\begin{equation}
    \tilde{v} \tilde{\xi}^{z-1} = v \xi^{z
-1}/D_{\alpha},
\end{equation} 
which effectively quantifies the extent to which the medium (whose relaxation rate is controlled by $\xi$, see \cref{eq:alpha_q}) is driven out of equilibrium. 
Hence, the asymptotic behavior for $\tilde{\xi} \gg 1$ is characterized by 
%the 
a
$\tilde{v}^{-2}$ scaling,
which is the same as that 
%typical 
of the scaling function $\mathcal{F}^{A/B}$ in the regime of large driving velocity, see the denominator of the integral in \cref{zeroT_rec_scale}. %\dav{OK?}%\textbf{[!!!]} \dav{What? (I suggest you to change color to your comments, or we'll forget to remove them before submission!}

%%%%%%%%%%%%%%%%%%%%%%%%%%%%%%%%%%%%%%%%%%%%
%%%%%%%%%%%%%%%%%%%%%%%%%%%%%%%%%%%%%%%%%%%%

\begin{figure}
    \centering
        \includegraphics[width=0.45\linewidth]{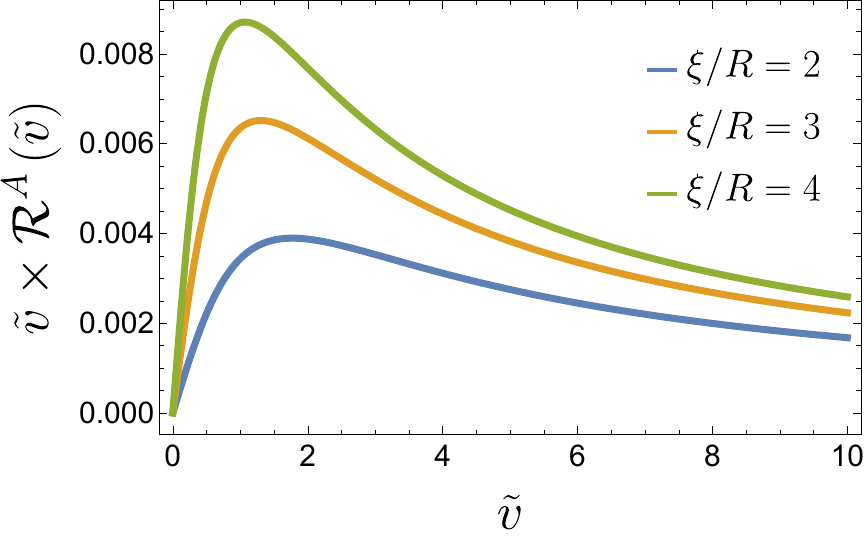}
          \put(-200,6){(a)}
          \hspace{10pt}
        \includegraphics[width=0.45\linewidth]{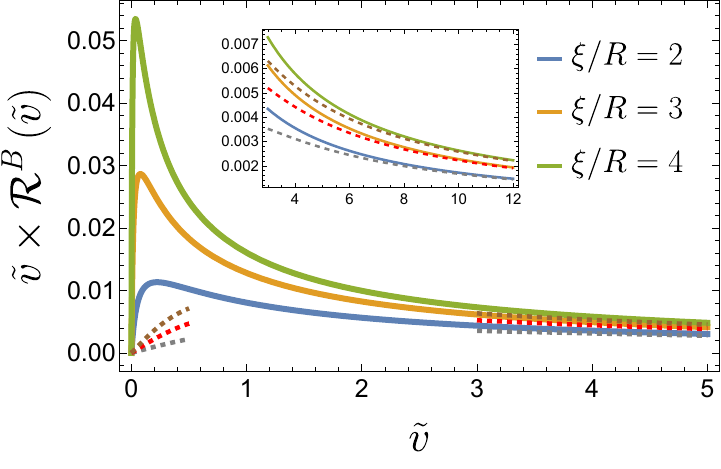}
          \put(-200,6){(b)}
\caption{%\ag{Please check, I modified this.} 
Dimensionless recoil range  $ \tilde v \times \mathcal{R}^{A/B}$ (see Eqs.~\eqref{zeroT_rec_scale_def} and~\eqref{eq:range_def}) as a function of the dimensionless driving velocity $\tilde v$ in spatial dimension $d=3$, for (a) model A or (b) model B, and various values of the correlation length $\xi$. 
In order to ease the comparison between these two dynamics, the dashed lines in panel (b) correspond to the behavior of the recoil range for model A at small or large velocities. As indicated in the inset, for large velocities, the curves corresponding to models A and B coincide.
}
\label{fig:vrange}
\end{figure}

The dimensionless recoil range $\tilde v \times \mathcal{R}^{A/B}$  
is plotted  as a function of the dimensionless velocity $\tilde{v}$ for $d=3$ in Fig.~\ref{fig:vrange}(a)
for model A. In particular, one observes that it vanishes at $\tilde v=0$  as long as the correlation length $\xi$ remains finite, while it reaches a finite value as  $\xi \to \infty$, according to Tab.~\ref{tab:small_v_scale}.  
Note that the value  $\tilde v_{\mathrm{max}}$ of the velocity $\tilde v$ at which $\tilde{v} \times \mathcal{R^A}$ 
is maximal decreases as $\tilde{\xi}$ increases.
In Fig.~\ref{fig:vrange}(b)
we plot $\tilde v \times \mathcal{R}^{B}(\tilde v)$ with solid lines; additionally, we plot $\tilde v  \times\mathcal{R}^{A}(\tilde v)$ with dashed lines to demonstrate that the dimensionless recoil range of the particle differs considerably between the cases of model~A and model~B dynamics for $\tilde{v}\ll 1$. We recall that $\tilde v$ --- defined in Eq.~\eqref{eq:Weissenberg} ---  describes the relative significance of the effects of the deformation compared with the relaxation of the field. 
Given that the expressions for the relaxation times depend on the dynamics of the field, as manifested by their $\alpha$-dependence in \cref{eq:alpha_q},
we expect the final position of the recoil to differ considerably in the relaxation-dominated ($\tilde{v} \ll 1$) regime. In fact, the long-time limit of the scaling function introduced in \cref{eq:range_def} in the limit of $v \to 0$ takes the form
\begin{equation}
    \mathcal{R}^{A/B}\left(1/\tilde \xi, \tilde v\right) \xrightarrow[\tilde v \to 0]{} \int \frac{\mathrm{d}^d\mathbf{q}}{(2 \pi )^d}\frac{q_1^2}{q^{2 \alpha} \left(q^2 + 1/\tilde{\xi}^2 \right)^{3}} |\hat{V}_{\mathbf{q}}|^2,
    \label{eq:v_small_R}
\end{equation}
which manifestly depends on the dynamics considered (via the factor $q^{2 \alpha}$ in the denominator\footnote{We recall that the integral in \cref{eq:v_small_R} may possibly exhibit infrared divergences, as already discussed after \cref{eq:crit_recoil_asympt}, and in Tab.~\ref{tab:small_v_scale}.}).  
This limit of weak driving out of equilibrium may be regarded as a linear response description of the recoil --- indeed, using \cref{zeroT_rec_scale_def,eq:range_def}, one deduces that $\expval*{ \mathbf{ \Delta X}^{(2)}(t \to \infty) } \propto v \mathcal{R}^{A/B}(1/\tilde \xi, \tilde v=0)$ for $v \to 0$, i.e., a linear dependence on $v$.
%(note the prefactor $g v \tau_R^2/\tau_X$ in \cref{zeroT_rec_scale_def}). 

Finally, in the opposite limit
%In the case of 
$\tilde{v} \gg 1$, the curves corresponding to model~A and model~B collapse onto each other
(this is better visible in the inset of 
Fig.~\ref{fig:vrange}(b)). Indeed, for large $\tilde{v}$, the expression for $\mathcal{R}^{A/B}$  (see Eqs.~\eqref{zeroT_rec_scale} and \eqref{eq:range_def}) %given in \cref{eq:range_def} 
becomes $\alpha$-independent:
\begin{equation}
\mathcal{R}^{A/B}\left(1/\tilde{\xi}, \tilde v \right) \sim  \frac{1}{\tilde{v}^2} \int\frac{\mathrm{d}^d  \mathbf{q}}{(2 \pi)^d}\frac{1}{q^2 + 1/\tilde{\xi}^2}|\hat{V}_{\mathbf{q}}|^2 , \qquad \mbox{for}\qquad \tilde{v}
\gg 1.
\end{equation}

%%%%%%%%%%%%%%%%%%%%%%%%%%%%%%%%%%%%%%%%
%%%%%%%%%%%%%%%%%%%%%%%%%%%%%%%%%%%%%%%%

\section{Non-equilibrium steady-state configuration of the field}
\label{sec:NESS}

Having discussed the properties of the recoil in the previous sections, below we turn our attention to the slow-relaxing field, since the force acting on the tracer is related to its gradient via Eq.~\eqref{eq:force_def}. 
In particular, we shall focus on the 
average field profile
in the non-equilibrium steady state that the system reaches at long times in the co-moving frame when the tracer is driven by the moving trap. 
The discussion below follows the one presented in Ref.~\cite{Venturelli_2023}, and builds upon the results reported therein and in Ref.~\cite{venturelli2023stochastic}. 
Beyond these references, we present here a detailed characterization of the steady-state configuration of the field for model A at criticality and for model B both at the critical point and in its vicinity. Such a characterization provides insight into the mechanism for the divergence of the recoil range and the algebraic asymptotic long-time  behavior of the scaling function reported in \cref{eq:crit_recoil_asympt}.

 We focus on the part of the protocol during which the particle is driven, i.e., on $t \leq 0$, and measure the particle position $\mathbf{X}$ from the minimum of the moving trap by defining $\mathbf{Z} = \mathbf{X} - \mathbf{v}t$. It is also convenient to introduce the translated field $\varphi(\mathbf{x}, t) = \phi(\mathbf{x} + \mathbf{v}t, t)$. Note that both $\mathcal{H}_{\phi}$ and $\mathcal{H}_{\mathrm{int}}$ given by Eqs.~\eqref{eq:gaussian_hamiltonian_rec} and \eqref{eq:Hint} are invariant under spatial translation, i.e., $\mathcal{H}_{\phi}[\phi] = \mathcal{H}_{\phi}[\varphi]$ and $\mathcal{H}_{\mathrm{int}}[\phi, \mathbf{X}] = \mathcal{H}_{\mathrm{int}}[\varphi, \mathbf{Z}]$. Moreover, the noise acting on the particle and the field is homogeneous in space. 
 Hence, the equations of motion \eqref{eq:field_four} and \eqref{eq:part} can be expressed in terms of $\mathbf{Z}$ and $\varphi_{\mathbf{q}}$ as~\cite{Venturelli_2023} 
 \begin{gather}
      \dot{\mathbf{Z}} = -\mathbf{v} - \kappa \nu \mathbf{Z} + \lambda \nu \int \frac{\mathrm{d}^d  \mathbf{q}}{(2 \pi)^d} i \mathbf{q}V_{\mathbf{-q}} \varphi_{\mathbf{q}}e^{i \mathbf{q} \cdot \mathbf{Z}} + \bm{\xi}, \label{eq:comove_part} \\ 
      \left[\partial_t + \alpha_q - i \mathbf{q}\cdot \mathbf{v} \right]\varphi_{\mathbf{q}}(t) = \lambda D_{\alpha} q^{\alpha}V_{\mathbf{q}}e^{-i\mathbf{q} \cdot \mathbf{Z}} + \eta_{\mathbf{q}},
      \label{eq:comove_field}
 \end{gather}
where $\alpha_q$ is defined in Eq.~\eqref{eq:alpha_q}, and $\varphi_{\mathbf{q}}(t)$ is the Fourier transform of $\varphi(\mathbf{x},t)$ (according to the definitions in \cref{eq:FT}). Note that changing the frame of reference essentially amounts to changing $\alpha_q \mapsto (\alpha_q - i\mathbf{q}\cdot \mathbf{v})$ on the left-hand side of the equation of motion of the field. 
 
For
a particle dragged through the medium for a sufficiently long time,
the system is expected to reach a stationary state characterized by $\langle \dot{\mathbf{Z}}\rangle_{\mathrm{ss}} = 0$ and $\langle \partial_t \varphi_{\mathbf{q}}\rangle_{\mathrm{ss}} =0$, which is satisfied for (see Eqs.~\eqref{eq:comove_part} and \eqref{eq:comove_field})
 \begin{gather}
     \langle \mathbf{Z}\rangle_{\mathrm{ss}} = - \mathbf{v}/(\kappa \nu) + \frac{\lambda}{\kappa}\int \frac{\mathrm{d}^d \mathbf{q}}{(2 \pi)^d} i \mathbf{q} V_{-\mathbf{q}}\langle \varphi_{\mathbf{q} }e^{i \mathbf{q}\cdot \mathbf{Z}}\rangle_{\mathrm{ss}},
     \label{eq:Z_NESS}\\[2mm]
     \langle \varphi_{\mathbf{q}}\rangle_{\mathrm{ss}} = \frac{\lambda D_{\alpha} q^{\alpha} V_{\mathbf{q}} \langle e^{- i \mathbf{q}\cdot \mathbf{Z}} \rangle_{\mathrm{ss}}}{\alpha_q - i\mathbf{q}\cdot \mathbf{v}},
     \label{eq:phi_NESS}
 \end{gather}
where $\langle \cdots\rangle_{\mathrm{ss}}$ indicates the expectation value taken in the stationary state. 
The fact that the system is driven out of equilibrium and that the probability density function in the stationary state is not known a priori (being generically different from the Boltzmann distribution), in conjunction with the coupling between the field $\varphi_{\mathbf{q}}$ and the particle position $\mathbf{Z}$, makes it very difficult to evaluate $\langle e^{- i \mathbf{q}\cdot \mathbf{Z}} \rangle_{\mathrm{ss}}$ and $\langle \varphi_{\mathbf{q} }e^{i \mathbf{q}\cdot \mathbf{Z}}\rangle_{\mathrm{ss}}$ on the right-hand side of Eqs.~\eqref{eq:Z_NESS} and \eqref{eq:phi_NESS}. 
Previous studies~\cite{Venturelli_2022,Basu_2022,Venturelli_2022_2parts,venturelli2023stochastic} considered a perturbative expansion of these expectation values in powers of $\lambda$ in order to derive predictions in 
the weak-coupling limit at the lowest non-vanishing order.  Qualitatively, assuming $\lambda V(\mathbf{x}) >0$, the resulting steady-state field profile $\left\langle  \varphi(\mathbf{x}) \right\rangle_{\mathrm{ss}}$, which is the inverse Fourier transform of $\left\langle  \varphi_{\mathbf{q}} \right\rangle_{\mathrm{ss}}$ in \cref{eq:phi_NESS}, is enhanced in the vicinity of the particle 
and stretched in the direction opposite to the dragging velocity (see \cref{fig:cartoon}). We will refer to this field configuration as the \textit{shadow}.  Its maximum 
turns out to be located behind the particle, and the presence of $\left\langle  \varphi(\mathbf{x}) \right\rangle_{\mathrm{ss}}$ induces an additional friction on the tracer, causing a shift of its steady-state position, as reported in Eq.~\eqref{eq:shift}.

\subsection{The noiseless limit}

In order to investigate the qualitative  features of the steady-state field configuration $\left\langle  \varphi(\mathbf{x}) \right\rangle_{\mathrm{ss}}$, it is convenient to neglect thermal noise by setting $T=0$. 
In this case, the dynamics in Eqs.~\eqref{eq:field_four} and \eqref{eq:part} becomes deterministic. Correspondingly, we denote $\langle \mathbf{Z}\rangle_{\mathrm{ss}} = \mathbf{Z}^{(\mathrm{ss})}$ and $\langle \varphi_{\mathbf{q}} \rangle_{\mathrm{ss}} = \varphi^{(\mathrm{ss})}_{\mathbf{q}}$, which satisfy $\langle \varphi_{\mathbf{q}} e^{i \mathbf{q} \cdot \mathbf{Z}}\rangle_{\mathrm{ss}} = \varphi_{\mathbf{q}}^{(\mathrm{ss})} e^{i \mathbf{q} \cdot \mathbf{Z}^{(\mathrm{ss})}}$ and $\langle e^{ - i \mathbf{q}\cdot \mathbf{Z}} \rangle_{\mathrm{ss}} = e^{ - i \mathbf{q}\cdot \mathbf{Z}^{(\mathrm{ss})}}$. Accordingly, Eqs.~\eqref{eq:Z_NESS} and \eqref{eq:phi_NESS} become 
 \begin{gather}
\mathbf{Z}^{(\mathrm{ss})} = - \mathbf{v}/ (\kappa \nu) -  
 \frac{\lambda^2}{ \kappa} \int \frac{\mathrm{d}^d    \mathbf q}{(2 \pi)^d}  \mathbf{q}D_{\alpha}q^{\alpha} |V_{\mathbf{q}}|^2\frac{\mathbf{q} \cdot\mathbf{v}}{\alpha_q^2 +  (\mathbf{q}\cdot \mathbf{v})^2}, 
 \label{eq:ZZ_NESS_T0} \\[2mm] 
 \varphi^{(\mathrm{ss})}_{\mathbf{q}} = \frac{\lambda D_{\alpha} q^{\alpha}V_{\mathbf{q}} e^{-i \mathbf{q} \cdot \mathbf{Z}^{(\mathrm{ss})}}}{\alpha_q - i\mathbf{q}\cdot \mathbf{v}}.
 \label{eq:Z_NESS_T0}
 \end{gather}
In passing, we note that Eq.~\eqref{eq:shift_T0} --- which describes the shift of the steady-state particle position obtained at $T=0$ in a perturbative expansion in
$\lambda$ --- coincides with the $\lambda$-dependent term in Eq.~\eqref{eq:ZZ_NESS_T0}, in which no assumption regarding the magnitude of $\lambda$ was made, yielding an exact expression. Accordingly, higher-order terms in the perturbative expansion are generated by thermal effects.

We note that, since the coordinate system is chosen such that the minimum of the optical trap is at the origin at time $t=0$, the steady-state configuration of the field provides the initial condition for the field dynamics for $t>0$, according to
  \begin{equation}
      \phi(\mathbf{x}, t=0) = \varphi^{(\mathrm{ss})}(\mathbf{x}).
      \label{eq-phi-ic}
  \end{equation}
 Moreover, as we discuss in, c.f., Eq.~\eqref{222}, the part  $\lambda\tilde \phi^{(1)}_{\mathbf{q}}(t)$ of the field contributing to the force~\eqref{eq:force_def} within the perturbative expansion in Eqs.~\eqref{eq:particle_expansion} and \eqref{eq:field_expansion} is  given by
\begin{equation}
    \lambda\tilde \phi^{(1)}_{\mathbf{q}}(t) = e^{- \alpha_q t} \varphi_{\mathbf{q}}^{(\mathrm{ss})},
    \label{eq:field_relax_forcing}
\end{equation}
for $t>0$ and in the noiseless limit $T=0$.
We conclude that it is the slow relaxation of $\lambda\tilde \phi^{(1)}_{\mathbf{q}}(t)$ that gives rise to a non-zero recoil: indeed, if the field relaxed instantaneously (corresponding to the formal limit $\alpha_q\to\infty$ taken in \cref{eq:field_relax_forcing}), then no systematic drift of the particle would be observed after its release from the trap. 
We note also  that plugging Eq.~\eqref{eq:Z_NESS_T0} into the expression for the field-induced force in Eq.~\eqref{eq:force_def} renders a time-independent force, from which the short-time expression for the recoil in Eq.~\eqref{eq:rec_smallt} with $T=0$ can be easily retrieved.

\subsection{The steady-state field configuration}
\label{sec:ss}

\begin{figure}
    \centering
    \begin{tabular}{cc}
        \includegraphics[width=0.45\linewidth]{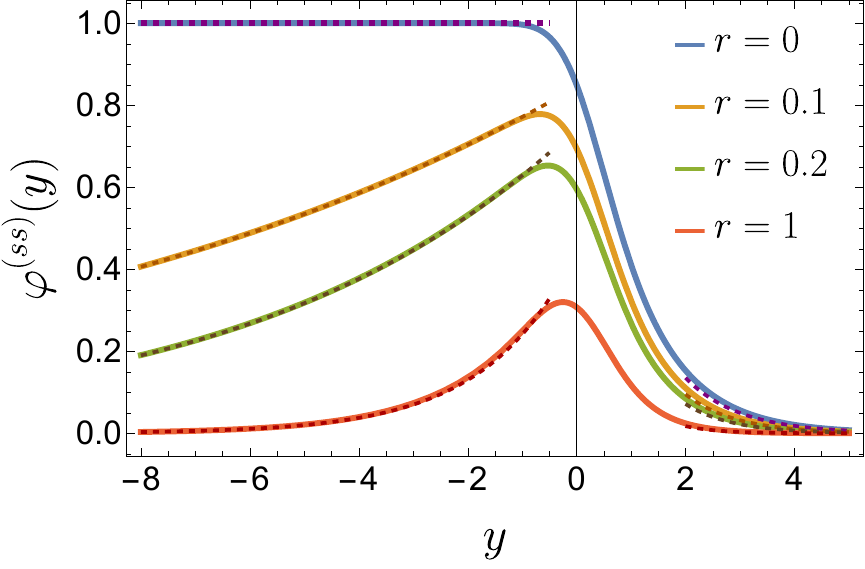}&
        \includegraphics[width=0.45\linewidth]{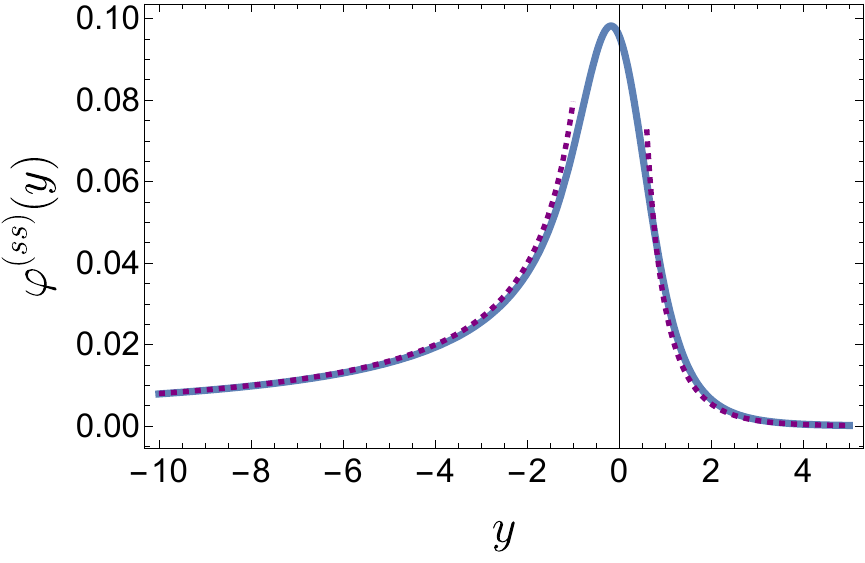}
        %\\[-2mm]
            %(a) & (b)
    \end{tabular}
    \put(-420,-55){(a)}
    \put(-205,-55){(b)}
\caption{Steady-state configuration of the field in the co-moving  reference frame for model A dynamics, as a function of the coordinate $y$ along the driving direction, and in the noiseless limit ($T=0$). Panel (a) corresponds to $d=1$ and  various values of $r$, while panel  (b) to $d=3$ at criticality ($r=0$).
In both panels $R = 0.5$, while $\lambda$, $D_0$, and $v$ are set to one.
The dashed lines in both panels correspond to the expressions in the case of a point-like particle, i.e., for $R\to 0$, reported in 
Eq.~\eqref{shadow_asympt}. Note that for $d=1$ at the critical point, the field configuration attains a non-vanishing value for $y\to  -\infty$.
For $d=3$, instead, the field profile decays as $\sim 1/|y|$ for $y\to -\infty$, in agreement with  Eq.~\eqref{eq:tail_power_law}. 
} 
\label{fig:shadow_A}
\end{figure}

At distances from the particle much larger than the particle size $R$,  
the field profile 
in the driven steady state can be described analytically by approximating the interaction kernel in Eq.~\eqref{eq:Hint} by a delta distribution which corresponds to the limit of a point-like particle, i.e., $R \to 0$.  
In particular, we focus on the behavior of the field configuration in the steady state as a function of the coordinate $y = \left(\mathbf{x} - \mathbf{Z}^{(\mathrm{ss})} \right) \cdot \mathbf{\hat{e}}_1$ along the dragging axis $\mathbf{\hat{e}}_1$, with $x_i = 0$ for $i \geq 2$ (where, as previously done, we assumed $\mathbf{v} = \mathbf{\hat{e}_1} v$).
 For model A, one obtains the field profile in the case of $V(\mathbf{x}) = \delta^{(d)}(\mathbf{x})$ in the form (see \ref{app:crit_tail_A} for details)
 \begin{equation}
     \varphi^{(\mathrm{ss})}(y) = \begin{cases}
         \frac{\lambda D_0}{v}\frac{1}{\sqrt{1 + (2D_0/v \xi)^2}}  \exp\left[-\frac{v}{2D_0}\left(|y|\sqrt{1 +(2D_0/v\xi)^2}  + y \right) \right], &\mbox{for}\quad d=1,  \\[3mm] 
         \frac{\lambda}{2\pi }K_0\left(\frac{|y| v}{2 D_0} \sqrt{1 +(2D_0/v\xi)^2} \right) \exp\left[-\frac{v}{2D_0}y\right] , &\mbox{for}\quad d=2,  \\[3mm]
         \frac{\lambda}{4 \pi |y|}\exp\left[-\frac{v}{2D_0}\left(|y|\sqrt{1 +(2D_0/v\xi)^2}  + y \right) \right], &\mbox{for}\quad d=3,
     \end{cases}
     \label{shadow_asympt}
 \end{equation}
 where $K_n(x)$ is the modified Bessel function of the second kind of order $n$. Equation~\eqref{shadow_asympt} allows us to define two length scales $l_+$ and $l_-$, describing the spatial decay of the tails of the shadow away from the particle along the direction of the driving or opposite to it, respectively, such that
 \begin{equation}
     \varphi^{(\mathrm{ss})}(y\to \pm \infty) \simeq   \exp({-|y|/l_\pm}).
 \end{equation}
 The inverse of these lengths is given by
 \begin{equation}
  (l_\pm)^{-1} = \frac{v}{2D_0}\left(\sqrt{1 +(2D_0/v \xi)^2}
 \pm 1\right),
 \label{eq:elplusmin}
 \end{equation}
 which coincide with the expressions reported in the supplemental material of Ref.~\cite{venturelli2023stochastic}.
 Note that, generically, $l_->l_+$, i.e., the field profile in front of the particle decreases more rapidly than that on its back. Moreover,
 upon approaching criticality, one has
  \begin{equation}
     l_+ \xrightarrow[\xi\to \infty]{} 
     %\frac{D_0}{v}
     D_0/v
      \quad \mbox{and} \quad  l_-  \xrightarrow[\xi \to \infty]{} +\infty. 
 \label{eq:lplus_crit}
 \end{equation}
For a particle with finite size, the maximum of the 
field profile is located to the left of the steady-state position of the particle, as shown in Fig.~\ref{fig:shadow_A}.

%%%%%%%%%%%%%%%%%%%%%%%%%%%%%%%%%%%%%%%%%%%%%%%%%%%%%%
%%%%%%%%%%%%%%%%%%%%%%%%%%%%%%%%%%%%%%%%%%%%%%%%%%%%%%

We note that, according to Eq.~\eqref{eq:elplusmin}, $l_+$ is always finite, and thus the shadow in front of the particle
decays exponentially for any value of $\xi$. 
On the contrary, at the critical point, the lengthscale $l_-$ describing the decay of the
field profile behind the particle diverges, and the profile exhibits an algebraic decay according to\footnote{Note that Eq.~\eqref{eq:tail_power_law} agrees with Eq.~\eqref{shadow_asympt}, since $e^xK_0(x) \sim \sqrt{\pi/(2x)}$ for large $x$. \label{foot}}
 \begin{equation}
     \varphi^{(\mathrm{ss})}(y \to -
     \infty)  \simeq \frac{\lambda}{(4 \pi |y|)^{(d-1)/2}} \left( \frac{v}{D_0} \right)^{(d-3)/2},
     \label{eq:tail_power_law}
 \end{equation}
as shown in~\ref{app:crit_tail}. 
Interestingly, in $d=1$, the 
field profile
approaches a finite value as $y \to - \infty$. 
This behavior can be traced back to the fact that a scalar Gaussian field theory is peculiar in $d\leq 2$ at the critical point, in that the long-distance behavior of its correlation function 
    \begin{equation}
        \langle \phi(\mathbf{x})\phi(0)\rangle \sim|\mathbf{x}|^{2-d}
        \label{eq:scalarFT}
    \end{equation}
is characterized by a linear growth of correlations in $d=1$~\cite{LeBellac}. 
Similarly, we attribute the divergence of the critical recoil range  reported in Eq.~\eqref{eq:range_asympt} to this fact. 

In passing, we note that 
in the limit in which the (non-critical) field equilibrates quickly as the particle moves (e.g., $D_0 \to \infty$), both length scales defined in Eq.~\eqref{eq:elplusmin} converge to the correlation length~$\xi$:
    \begin{equation}
        l_{\pm} \rightarrow \xi \quad\mbox{for} \quad D_0 \to \infty.
    \end{equation}
Accordingly, in this \textit{adiabatic} limit, the shape of the 
field
profile becomes independent of the dragging velocity and, in the co-moving frame, it is the same as the equilibrium one at rest. Since the mobility of the field enters Eq.~\eqref{eq:elplusmin} only via the combination $v/D_0$, the adiabatic limit is actually equivalent to the limit $v \to 0$ with fixed $\xi$ and $D_0$. 

In the case of model B, determining the features of the 
field
profile is technically more demanding, and thus we postpone the discussion of the non-critical case to \ref{app:crit_tail}. Here, we report the critical behavior and the qualitative features for finite correlation lengths.   
At the critical point, for $d=1$ and for the delta-like interaction potential, the stationary 
field
profile 
$\varphi^{(\mathrm{ss})}(y)$ 
is given by
    \begin{equation}
        \varphi^{(\mathrm{ss})}(y) = \begin{cases}
            \frac{\lambda D_2^{1/3}}{3v^{1/3}} e^{-|y|\left(v/D_2\right)^{1/3} }, & \quad \mbox{for}  \quad y<0, \\ 
            \frac{2\lambda D_2^{1/3}}{3v^{1/3}} e^{-\frac{y}{2}\left(v/D_2\right)^{1/3} }\sin \left( \frac{\pi}{6} - \frac{\sqrt{3}}{2}y(v/D_2)^{1/3}\right), & \quad \mbox{for}  \quad y>0. 
        \end{cases}
        \label{eq:shadow_tail_B}
    \end{equation}
We notice that, 
since the dynamics conserves locally the field,
the integral over the whole space of the steady-state field configuration is constant;
in particular, here it vanishes, because the dynamics conserves the $\phi_{\mathbf{q} =0}$ mode which  in the calculation we had set to zero at $t_0 \to - \infty$. 
In spatial dimension $d=3$ and in the limit of a point-like particle, the field profile in the direction of the driving force is given by
    \begin{equation}
        \varphi^{(\mathrm{ss})}(y) = \begin{cases}
            \frac{\lambda }{8 \pi |y|}\left(1 + e^{-|y|\left(v/D_2\right)^{1/3} }\right), & \quad\mbox{for}\quad y<0, \\[2mm] 
            \frac{\lambda }{4 \pi y } e^{-\frac{y}{2}\left(v/D_2\right)^{1/3} }\cos \left( \frac{\sqrt{3}}{2}y(v/D_2)^{1/3}\right), & \quad\mbox{for}\quad y>0. 
        \end{cases}
        \label{eq:shadow_B_asympt}
    \end{equation}
Accordingly, at criticality, a qualitative difference between model A and model B is observed in the behavior of $\varphi^{(\mathrm{ss})}(y)$  for $|y| \gg R$. In fact, for model A, the order parameter decays monotonically upon increasing the distance from the particle, as in Eq.~\eqref{shadow_asympt}. For model B, instead, the field profile in the direction opposite to the drive still decreases monotonically, while in the direction of the drive it displays damped oscillations, as in Eq.~\eqref{eq:shadow_B_asympt} above. 
Note that, according to Eqs.~\eqref{eq:lplus_crit} and \eqref{eq:shadow_tail_B}, when the limits  $R \to 0$ and $\xi \to \infty$ are taken simultaneously --- i.e., when one considers a point-like particle in a critical medium --- the field profile varies in space on the lengthscale defined in \cref{eq:xi-v}, %\dav{it has already been defined in ! Moreover, I suggest to choose another name (i.e.~avoid $\xi$, see my comment above)}
 %\begin{equation}%
%     l_{v, \alpha} = \left( \frac{D_{\alpha}}{v}\right)^{1/(\alpha + 1)},
  %   \label{eq:xi_v_def}
 %\end{equation}
which describes the relative significance of the effects due to the drive and the relaxation of the medium. In particular, this ratio is related to the Weissenberg number $\mathrm{Wi}$ introduced in Eq.~\eqref{eq:Weissenberg} via $2\mathrm{Wi} = \left( R/l_{v, \alpha}\right)^{\alpha +1}$.

 \begin{figure}
    \centering
            \includegraphics[width=0.45\linewidth]{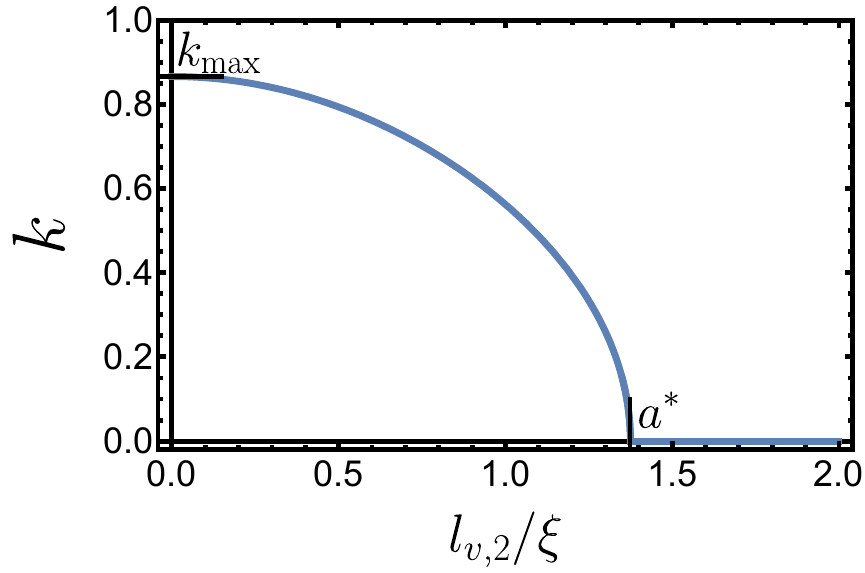} 
\caption{Dependence of the dimensionless wavevector $k$  of the damped  oscillations of the field profile in the direction of the drag (see Eq.~\eqref{eq:front_beh}), for model~B and a delta-like interaction kernel, defined in 
\cref{eq:front_beh} as a function of $a \equiv l_{v,2}/\xi$ (see Eq.~\eqref{eq:def-a-v}). 
Spatial oscillations are observed for $l_{v,2}/\xi<a^*$ (see Eq.~\eqref{eq:condition}), and the wavevector $k$ attains its maximal value $k_{\mathrm{max}}  = \sqrt{3}/2$ in the limit $l_{v,2}/\xi \to 0$. 
When $l_{v,2}/\xi$ approaches the value $a^*$ from below, the wavevector $k$ vanishes according to $ k \sim \sqrt{a^* - \left(l_{v,2}/\xi\right)}$. 
%\dav{I hate to say that, but we have to update the x-label of the plot...} \textbf{[good call - I am on it!]}
}
\label{fig:freq}
\end{figure} 

In the case of a non-critical medium governed by model~B dynamics, the steady-state field configuration tuns out to display damped oscillations away from the particle in the direction of the drive  
%only provided that 
if the velocity $v$ is sufficiently large, i.e., for 
\begin{equation}
    \frac{l_{v,2}}{\xi} < \frac{3^{1/2}}{2^{1/3}} \equiv a^* \approx 1.37,
    \label{eq:condition}
\end{equation}
irrespective of the spatial dimensionality of the system. 
In particular, the behavior of the field for $y>0$ in the case of model~B, in $d=1$, and in the limit of a point-like particle, as a function of 
\begin{equation}
a \equiv {l_{v,2}}/{\xi},
\label{eq:def-a-v}
\end{equation}
can be written as
\begin{equation}
    \varphi^{(\mathrm{ss})}(y) \propto \begin{cases}
            \exp\left[- \gamma(a) y/l_{v,2}  \right] \cos\left[ k(a) y/l_{v,2} - \psi(a)\right], & \quad\mbox{for} \quad a < a^*, \\[2mm] 
            \exp\left[- \gamma_1(a) y/l_{v,2}  \right]  - A(a)\exp\left[- \gamma_2(a)  y/l_{v,2}  \right], & \quad\mbox{for} \quad a > a^*,
        \end{cases} 
        \label{eq:front_beh}
\end{equation}
where $1>A(a)>0$, $\gamma(a)>0$,  
 $\gamma_1(a)>\gamma_2(a)>0$, and $k(a)$ is a dimensionless wavevector to be discussed below. 
Accordingly, for ${l_{v,2}}/{\xi} > a^*$, the field profile is described by a superposition of two exponential functions with amplitudes of opposite sign. Notably, the function with the negative amplitude decays more slowly than the other upon increasing the distance from the particle.  Explicit expressions for the field profile are presented in 
\ref{app:crit_tail_B}. 
The dimensionless wavevector $k$ introduced in \cref{eq:front_beh} is plotted as a function of $a$ (see Eq.~\eqref{eq:def-a-v}) in Fig.~\ref{fig:freq}. 
In the case of $d>1$, the transition from the damped-oscillations regime to that described by a superposition of two exponential functions also occurs at $a = a^*$ (see Eq.~\eqref{eq:condition}).% and the description ofthe front part of the field profile is analogous, up to algebraic prefactors in $y/l_{v,2}$.

%%%%%%%%%%%%%%%%%%%%%%%%%%%%%%%%%%%%%

\begin{figure}
    \centering
    \begin{tabular}{cc}
            \includegraphics[width=0.45\linewidth]{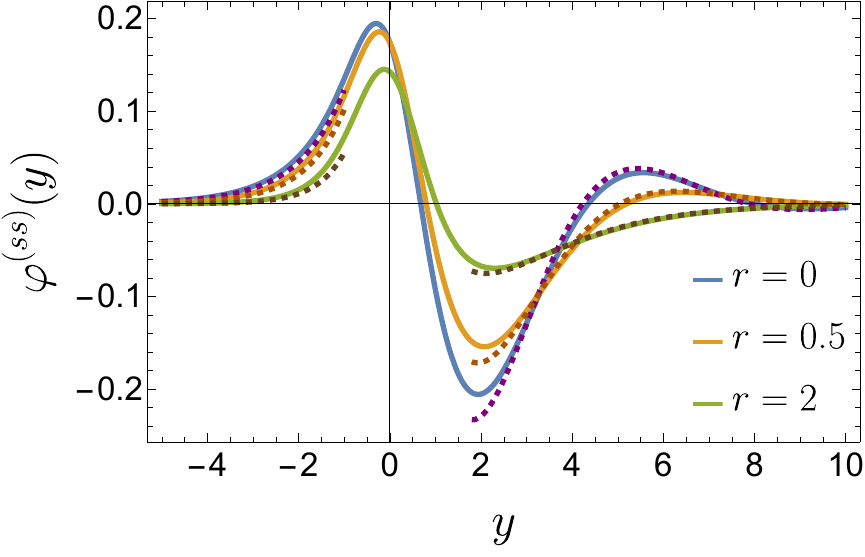} &
            \includegraphics[width=0.45\linewidth]{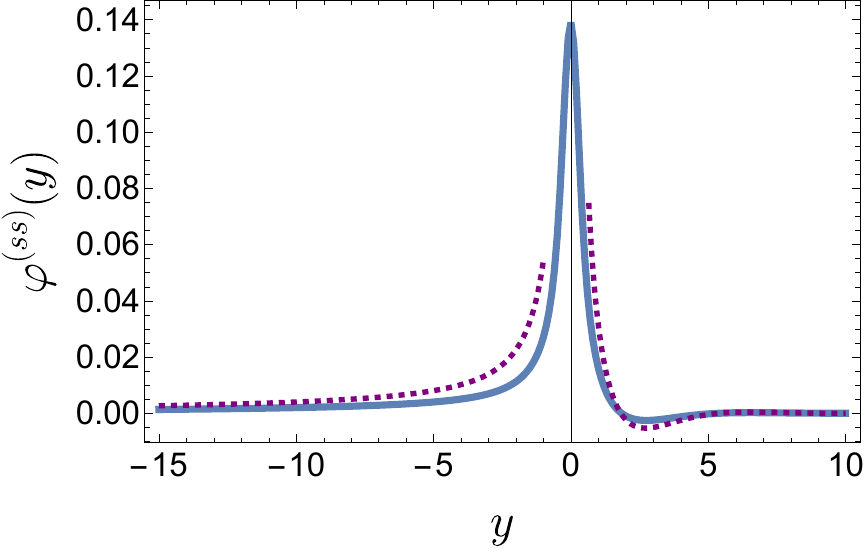} 
            %\\[-2mm]
        %(a) & (b) 
\end{tabular}
    \put(-420,-55){(a)}
    \put(-205,-55){(b)}
\caption{Steady-state configuration $\varphi^{(ss)}(y)$ of the field in the co-moving  reference frame for model~B, as a function of the coordinate $y$ along the driving direction, and in the noiseless limit ($T=0$).
Panel (a) corresponds to $d=1$ with $R = 0.5$ and various values of $r$,
wile panel (b) to $d = 3$  with $R=0.2$ and at criticality.
In both panels $\lambda$, $D_0$, and $v$ are set to one.
The dashed lines in both panels correspond to the expressions in the case of a point-like particle, i.e., for $R\to 0$, reported in Eqs.~\eqref{eq:app_shadow_b_2}--\eqref{eq:app_shadow_b_1} and \eqref{eq:shadow_B_asympt}.
Note that for $d=1$, the field configuration is characterized by overdamped oscillations for $r = 0$ and~$0.5$, and by a superposition of two exponential functions of opposite signs for $r = 2$.
These different qualitative behaviors are found in agreement with the condition for the presence of oscillations given in \cref{eq:condition}. 
}
\label{fig:shadow_B}
\end{figure} 
%%%%%%%%%%%%%%%%%%%%%%%%%%%%%%%%%%%%%
%%

Representative plots of the field profiles for model A and B are reported in Figs.~\ref{fig:shadow_A} and \ref{fig:shadow_B}, respectively, which assume the Gaussian interaction kernel reported in Eq.~\eqref{eq:Gauss_kernel}.
In particular, in panels (a) of both plots, the one-dimensional case 
is reported with various values of the parameter $r$, which controls the correlation length of the medium. The dashed lines in the plots correspond to the expressions in the $R \to 0$ limit, given in Eq.~\eqref{shadow_asympt} and 
\cref{eq:app_shadow_b_1,eq:app_shadow_b_2},
respectively. 
The plots of the steady-state field profile along the driving direction
in spatial dimension $d=3$ at criticality are reported in panels~(b) of 
Figs.~\ref{fig:shadow_A} and \ref{fig:shadow_B}, together with the corresponding asymptotic behaviors given in  Eqs.~\eqref{shadow_asympt}  and \eqref{eq:shadow_B_asympt}. 

Finally, we note that in $d=3$, irrespective of the choice of the dynamics of the field and of the interaction kernel, taking the limit $\xi \to \infty$ and subsequently $v \to 0$ yields, for the steady-state field profile along the direction of the driving, 
\begin{equation}
        \varphi^{(\mathrm{ss})}(y \to \pm\infty) \simeq \frac{\lambda}{4 \pi |y|},
        \label{eq:shadow_crit_nov}
\end{equation}
which, as expected, corresponds to the equilibrium configuration of the field with a point-like source. This indicates that in spatial dimension $d=3$, the limits $\xi \to \infty$ and $v \to 0$ commute, which is not the case for $d=1$ or 2. Again, the emergence of this issue can be traced back to the previously discussed peculiarity of the scalar Gaussian field theory at the critical point, see \cref{eq:scalarFT}.

\section{Conclusions}
\label{sec:conclusions}

In this work we described the dynamics of
an overdamped Brownian particle coupled to a fluctuating scalar Gaussian field, characterized by a correlation length $\xi$. The field is assumed to have a relaxational dynamics, with the possible presence of a local conservation law.   
The particle and the field are linearly coupled to each other, with the strength of their interaction determined by the coupling constant $\lambda$, and both are in contact with a thermal bath at temperature $T$. At times $t<0$, the particle is driven across the medium by a harmonic trap moving with constant velocity $\mathbf{v}$, and it is released from the trap at $t=0$. 
In describing the effective dynamics of the tracer particle, which is non-linear and non-Markovian, we resorted to a perturbative expansion in the coupling~$\lambda$. 
We demonstrated that, after the trap is turned off at time $t=0$, the tracer exhibits \textit{recoil}, i.e., it moves in the direction opposite to that of the dragging velocity. 

After deriving an expression for the average recoil in Eq.~\eqref{eq:recoil_full}, we focused on the noiseless limit corresponding to $T=0$, for which, in Eq.~\eqref{zeroT_rec_scale_def},  we introduced a scaling function  $\mathcal{F}^{A/B}\left(R/\xi, t/\tau_R, v \tau_R/R \right)$ --- where $\tau_R$ is the typical relaxation time of a critical field over a distance of the order of the effective particle size $R$. 
For small velocities~$v$, the amplitude of the final recoil is $\propto v$ as long as $t/\tau_R$ and $\xi/R$ remain finite. The non-trivial small-$v$ behavior arising for $\xi \to \infty$ and $t \to \infty$ is summarized in Tab.~\ref{tab:small_v_scale}.
We showed that, generically, the recoil eventually vanishes upon increasing $v$, due to an effective decoupling of the particle from the field.
In the critical limit $\xi \to \infty$, the eventual recoil has a finite amplitude for $d >2$, and it reaches its maximal value following an algebraic law $\sim t^{(2-d)/z}$, where $z$ is the dynamical critical exponent of the field, as plotted for $d=3$ in Figs.~\ref{fig:recoil-range}(a) and \ref{fig:recoil-rangeB}(a) for dissipative (model~A) and conservative (model~B) dynamics, respectively. 
For $d \leq 2$, as the system approaches its critical point, we described the divergence of the recoil range $\propto \xi^{2-d}$. Conversely, in the non-critical case, the recoil has a finite range for any spatial dimensionality $d$. For non-conserved dynamics, the final value of the recoil is approached exponentially in time,  whereas for conserved dynamics, the approach is algebraic $\propto t^{-d/2}$. A comprehensive summary of the long-time behavior of the scaling function $\mathcal{F}^{A/B}$ is reported in Tab.~\ref{tab:range}.

As it drives the initial evolution of the recoil, we investigated the spatial configuration of the field which develops around the particle in the steady state generated by the driving at time $t<0$, considering it in the reference frame co-moving with the trap. 
In the case of non-conserved dynamics and at $T=0$, two lengthscales $l_\pm$ naturally emerge, which describe the rate of the exponential decay towards zero  of the field  upon moving away from the particle in the direction ($l_+$) or opposite ($l_-$) to the driving, i.e., in front or behind the dragged particle. In the critical case $\xi \to \infty$, $l_+$ remains finite whereas $l_-$ diverges, and the decay of the field profile is characterized by an algebraic law $\propto 1/|y|^{(d-1)/2}$, where $y$ is the distance %from the center of the particle 
along the axis of the drive from the steady-state position of the tracer in the co-moving frame (see Eq.~\eqref{eq:tail_power_law}). We noted that, in the adiabatic limit in which the field equilibrates instantaneously around the current position of the particle, both $l_+$ and $l_-$ reduce to the correlation length $\xi$. In the case of conserved dynamics, depending on the values of the various parameters, 
it is possible to observe damped spatial oscillations 
of the average field profile in front of the dragged particle, whose wavevector is plotted in Fig.~\ref{fig:freq}.

This work may motivate an experimental study of recoil of colloidal particles dragged through a near-critical binary mixture, in a manner similar to the investigations reported in, e.g., Ref.~\cite{Cao_2023} for viscoelastic media. 
Colloidal particles in critical media are, by now, easily accessible experimentally \cite{Hertlein_2008, Gambassi_2009_exp, Paladugu_2016, Martinez_Entropy_2017, Magazzu_2019}, and the values of the correlation length that can be reached in these experiments are of the order of hundreds of nanometers, 
which is comparable
with the typical size of the colloidal particles.  Also, in experiments in viscoelastic media \cite{Gomez-Solano_2015}, the dragging velocities considered spanned the range between $0.4\mu m/s$ and $ 40 \mu m/s$.
In these experimental conditions, it should be possible to test, at least qualitatively, the behavior of the recoil predicted in this work.
However, a more quantitative description would require accounting for the effect of hydrodynamics and of the self-interaction $\propto \phi^4$ of the fluctuating order-parameter field. These issues are left for future investigations.

Finally, the framework adopted in this work could be adapted to describe other phenomena exhibited by a tracer particle coupled to a slowly relaxing structured medium ---
such as the \textit{Magnus effect} recently observed with rotating particles in viscoelastic media~\cite{Cao_2023}. 
This possibility calls for an extension of the model discussed here in order to describe a particle and a medium that are subject not only to a translational, but also to a rotational shear. Moreover, the discussion may be extended \rev{to systems in which the fluctuation-dissipation relation is not satisfied, e.g.,} by considering an active field, modeling a non-equilibrium bath of active particles (as investigated, e.g., in Refs.~\cite{Maes_2020, Granek_2022, Valadez_2023, Santra_2023, Maes_2025}), \rev{or conversely by considering an active particle evolving in contact with a passive field}.

 \section*{Acknowledgments}
 We thank 
 Clemens Bechinger, 
 Vincent D\'emery, Aljaz Godec, Matthias Kr\"uger,  Christian Maes, and Gennaro Tucci for stimulating discussions.
 AG acknowledges support from MIUR PRIN project “Coarse-grained description for non-equilibrium systems and transport phenomena (CO-NEST)” n.~201798CZL.
\appendix
\addtocontents{toc}{\fixappendix}

\section{Perturbative calculation of the correction to the particle's position}
\label{app:pert_cal}

In this Appendix we present
%, in a perturbative expansion,  
the derivation of the correction to the particle position $\mathbf{X}$ due to the interaction \eqref{eq:force_def} with the field. In particular, in \ref{app:pos_shift}  this is done while the particle is driven by the trap, resulting in the time-independent shift reported in Eq.~\eqref{eq:shift}, and in \ref{app-recoil} after the particle has been released from the confinement, yielding the recoil reported in Eq.~\eqref{eq:recoil_full}.

\subsection{Driven particle}
\label{app:pos_shift}
For $t\leq 0$, the particle is dragged by the moving  trap. 
As we are interested in the steady state attained at long times in the co-moving frame of the particle, we formally take the limit
$t_0 \to -\infty$. When the particle and the field are decoupled (i.e., with $\lambda=0$), according to 
\cref{eq:field_four,eq:part},
their position and configuration are given by  
\begin{align}
    \mathbf{X}^{(0)}(t) &= \mathbf{v}\left(t - \frac{1}{\nu \kappa}\right) + \mathbf{O}(t),
    \label{part_decoupled}
     \\ 
     \phi^{(0)}_{\mathbf{q}}(t) &= \int_{-\infty}^t\mathrm{d}t'e^{-\alpha_q(t - t')} \eta_{\mathbf{q}}(t'),
     \label{field_decoupled}
\end{align}
where 
$\mathbf{O}(t) = \int_{-\infty}^{t}\mathrm{d}t'\, e^{-\nu \kappa (t -t')}\bm{\xi}(t')$ is the Ornstein-Uhlenbeck process. This Gaussian process is characterized by a vanishing mean and two-point correlator~\cite{Tauber_2014}
    \begin{equation}
        \langle \mathrm{O}_i(t_1)\mathrm{O}_j(t_2) \rangle = \frac{T}{\kappa}e^{- \nu \kappa|t_2 - t_1
        |}\delta_{i,j}.
    \end{equation}
    Note that, since the initial conditions are imposed at $t_0 \to - \infty$, they are irrelevant in determining the steady-state configuration of the field in Eq.~\eqref{field_decoupled}. Moreover, in Eq.~\eqref{part_decoupled}, 
    we chose the coordinate system such that the minimum of the trap is at the origin at $t=0$.
    
     Using the expressions for the decoupled case (see Eqs.~\eqref{part_decoupled} and \eqref{field_decoupled}), we solve Eqs.~\eqref{eq:field_four} and  \eqref{eq:part} term-by-term within the perturbative expansion in $\lambda$ introduced in Eqs.~\eqref{eq:particle_expansion} and \eqref{eq:field_expansion}. 
     At the first order in $\lambda$, we obtain
    \begin{align}
    \mathbf{X}^{(1)}(t) &= i \nu \int \frac{\mathrm{d}^d   \mathbf{q}}{(2 \pi)^d}\mathbf{q}V_{-\mathbf{q}} \int_{-\infty}^t \!\!\mathrm{d}t' \, e^{-\nu \kappa(t - t') + i\mathbf{q}\cdot\left[\mathbf{v}\left(t'- \frac{1}{\nu \kappa}\right) + \mathbf{O}(t')\right]} \phi_{\mathbf{q}}^{(0)}(t'),
     \\ 
     \phi^{(1)}_{\mathbf{q}}(t) &= D_{\alpha}q^{\alpha}V_{\mathbf{q}} \int_{-\infty}^t \!\!\mathrm{d}t'\, e^{-\alpha_q(t - t') - i\mathbf{q}\cdot\left[\mathbf{v}\left(t'- \frac{1}{\nu \kappa}\right) + \mathbf{O}(t')\right]}.
     \label{field_t_neg}
\end{align}
At the second order in $\lambda$, Eq.~\eqref{eq:part} for the particle position 
renders the expression reported in \cref{eq:2nd_order_part} in the main text. 
We evaluate the average of each term therein separately, and write
\begin{align}
     &i \nu \int \frac{\mathrm{d}^d   \mathbf{q}}{(2\pi)^d} \mathbf{q}  V_{-\mathbf{q}} \left\langle e^{i \mathbf{q}\cdot \mathbf{X}^{(0)}(t)}\phi_{\mathbf{q}}^{(1)}(t) \right \rangle \n  \\ 
     &=i \nu \int \frac{\mathrm{d}^d   \mathbf{q}}{(2\pi)^d} |V_{\mathbf{q}}|^2D_{\alpha}q^{\alpha} \mathbf{q}\int_{-\infty}^t\mathrm{d}t'e^{-\alpha_q(t - t') + i\mathbf{q}\cdot \mathbf{v}(t - t')}\left\langle e^{i\mathbf{q
    }\cdot [\mathbf{O}(t) - \mathbf{O}(t')]}\right \rangle.
\end{align}
We now recall the property of the Ornstein-Uhlenbeck process (see, e.g., Appendix~C.1 of Ref.~\cite{Basu_2022}),
\begin{equation}
    \left\langle e^{i\mathbf{q
    }\cdot [\mathbf{O}(t) - \mathbf{O}(t')]}\right \rangle = e^{-q^2(T/\kappa)\left[1 - \exp(-\nu \kappa|t - t'|) \right]},
    \label{eq:app_OU_SF}
\end{equation}
and arrive at
\begin{align}
     &i \nu \int \frac{\mathrm{d}^d   \mathbf{q}}{(2\pi)^d} \left\langle\mathbf{q}e^{i \mathbf{q}\cdot \mathbf{X}^{(0)}(t)}V_{-\mathbf{q}}\phi_{\mathbf{q}}^{(1)}(t) \right \rangle  \n \\
     &=i \nu \int \frac{\mathrm{d}^d   \mathbf{q}}{(2\pi)^d} |V_{\mathbf{q}}|^2\frac{\alpha_q}{q^2 + r} \mathbf{q}\int_{-\infty}^t\mathrm{d}t'e^{-\alpha_q(t - t') + i\mathbf{q}\cdot \mathbf{v}(t - t')-q^2({T}/{\kappa})\left[1 - e^{-\nu \kappa(t - t')} \right] }.
     \label{eq:app_shift_term_1}
\end{align}
For the second term in 
\cref{eq:2nd_order_part},
we first average over the noise $\eta$ and find
\begin{align}
     &i \nu \int \frac{\mathrm{d}^d   \mathbf{q}}{(2\pi)^d} \mathbf{q}e^{i \mathbf{q}\cdot \mathbf{X}^{(0)}(t)}V_{-\mathbf{q}}i\mathbf{q}\cdot \left\langle\mathbf{X}^{(1)}(t)  \phi_{\mathbf{q}}^{(0)}(t) \right \rangle_{\eta} =-i \nu^2 \int \frac{\mathrm{d}^d   \mathbf{q}}{(2\pi)^d}\int \frac{\mathrm{d}^d   \mathbf{q}'}{(2\pi)^d} \mathbf{q}V_{-\mathbf{q}}V_{-\mathbf{q'}}(\mathbf{q}\cdot\mathbf{q'}) \n\\ 
     &\quad\times \int_{-\infty}^t \mathrm{d}t'e^{-\nu \kappa(t - t') +  i \mathbf{q}\cdot \left[\mathbf{v}\left(t - \frac{1}{\nu \kappa}\right) + \mathbf{O}(t)\right] + i \mathbf{q'}\cdot \left[\mathbf{v}\left(t' - \frac{1}{\nu \kappa}\right) + \mathbf{O}(t')\right]} \left\langle \phi^{(0)}_{\mathbf{q}} (t)\phi^{(0)}_{\mathbf{q'}} (t')\right\rangle_{\eta} \n \\ 
     &= i \nu^2 \int \frac{\mathrm{d}^d   \mathbf{q}}{(2\pi)^d} \mathbf{q}|V_{\mathbf{q}}|^2q^2 \frac{T}{q^2 + r} \int_{-\infty}^t \mathrm{d}t'e^{-\alpha_q(t - t')-\nu \kappa(t - t') +  i \mathbf{q}\cdot \left[\mathbf{v}\left(t - t'\right) + \mathbf{O}(t)- \mathbf{O}(t')\right] }, 
\end{align}
where we used the correlator of the Fourier transform of the field's noise in \cref{eq:noise_corr_ft}
%\begin{equation}
 %   \langle \eta_{\mathbf{q}}(t)\eta_{\mathbf{q'}}(t') \rangle_{\eta} = 2TD_{\alpha}q^{\alpha}(2 \pi)^d \delta^{(d)}(\mathbf{q} +\mathbf{q'}) \delta (t - t')
%\end{equation}
to derive, in the long-time limit~\cite{Tauber_2014},
\begin{equation}
    \left\langle \phi_{\mathbf{q}}^{(0)}(t)\phi_{\mathbf{q'}}^{(0)}(t') \right\rangle_{\eta} =  (2 \pi)^d\delta^{(d)}(\mathbf{q} + \,\mathbf{q'})\frac{T}{q^2 + r}e^{-\alpha_q|t - t'|}.
\end{equation}
Using again the property \eqref{eq:app_OU_SF} of the Ornstein-Uhlenbeck process, we arrive at
\begin{align}
     &i \nu \int \frac{\mathrm{d}^d   \mathbf{q}}{(2\pi)^d} \mathbf{q} V_{-\mathbf{q}} \left\langle e^{i \mathbf{q}\cdot  \mathbf{X}^{(0)}(t)}i\mathbf{q}\cdot \mathbf{X}^{(1)}(t)  \phi_{\mathbf{q}}^{(0)}(t) \right \rangle   \\ 
     %&=i \nu^2 \int \frac{\mathrm{d}^d   \mathbf{q}}{(2\pi)^d} \mathbf{q}|V_{\mathbf{q}}|^2q^2 \frac{T}{q^2 + r} \int_{-\infty}^t \mathrm{d}t'e^{-\alpha_q(t - t')-\nu \kappa(t - t') +  i \mathbf{q}\cdot \mathbf{v}\left(t - t'\right)}\left\langle e^{i\mathbf{q}\cdot (\mathbf{O}(t) - \mathbf{O}(t') )}\right\rangle \n \\ 
     &=i \nu^2 \int \frac{\mathrm{d}^d   \mathbf{q}}{(2\pi)^d} \mathbf{q}|V_{\mathbf{q}}|^2q^2 \frac{T}{q^2 + r} \int_{-\infty}^t \mathrm{d}t'e^{-\alpha_q(t - t')-\nu \kappa(t - t') +  i \mathbf{q}\cdot \mathbf{v}\left(t - t'\right) - q^2({T}/{\kappa})\left[1- e^{-\nu \kappa (t - t')} \right]}. \n
\end{align}
%\ag{I cut a line from the expression above....}
Combining this with Eq.~(\ref{eq:app_shift_term_1})
allows us to write
\begin{align}    
     &i \nu \int \frac{\mathrm{d}^d   \mathbf{q}}{(2\pi)^d} \mathbf{q} \left\langle e^{i \mathbf{q}\cdot \mathbf{X}^{(0)}(t)}V_{-\mathbf{q}}\left[\phi_{\mathbf{q}}^{(1)}(t)  + i\mathbf{q}\cdot \mathbf{X}^{(1)}(t)  \phi_{\mathbf{q}}^{(0)}(t)\right] \right \rangle  \n\\ &=i \nu \int \frac{\mathrm{d}^d   \mathbf{q}}{(2 \pi)^d}|V_{\mathbf{q}}|^2 \mathbf{q}\frac{1}{q^2 +r} \int_0^{\infty}\mathrm{d}s \, e^{-\alpha_q s - q^2({T}/{\kappa})\left(1 - e^{-\nu \kappa s }\right)}\left[\alpha_q + q^2 \nu T e^{-\nu \kappa s} \right]e^{i\mathbf{q}\cdot \mathbf{v} s}.
\label{eq:app-xx}
\end{align}
Noting that  
\begin{equation}
    \frac{\mathrm{d}}{\mathrm{d} s}e^{-\alpha_q s - q^2({T}/{\kappa})\left(1 - e^{-\nu \kappa s }\right)} = -\left[\alpha_q + q^2 \nu T e^{-\nu \kappa s} \right]e^{-\alpha_q s - q^2({T}/{\kappa})\left(1 - e^{-\nu \kappa s }\right)},
\end{equation} 
the integral over $s$ in Eq.~\eqref{eq:app-xx} can be computed by parts. The 
%
%\ag{I SHORTENED THE PRESENTATION HERE, by cutting some equations...}
%
%\begin{align}    
%     &i \nu \int \frac{\mathrm{d}^d   \mathbf{q}}{(2\pi)^d} \mathbf{q} \left\langle e^{i \mathbf{q}\cdot \mathbf{X}^{(0)}(t)}V_{-\mathbf{q}}\left[\phi_{\mathbf{q}}^{(1)}(t)  + i\mathbf{q}\cdot \mathbf{X}^{(1)}(t)  \phi_{\mathbf{q}}^{(0)}(t)\right] \right \rangle \n \\ &= -i \nu \int \frac{\mathrm{d}^d   \mathbf{q}}{(2 \pi)^d}|V_{\mathbf{q}}|^2 \mathbf{q}\frac{1}{q^2 +r} \int_0^{\infty}\mathrm{d}s \frac{\mathrm{d}}{\mathrm{d}s}\left[ e^{-\alpha_q s - q^2\frac{T}{\kappa}\left(1 - e^{-\nu \kappa s }\right)}\right ] e^{i\mathbf{q}\cdot \mathbf{v} s}.
%\end{align}
%We now integrate by parts and notice that the 
boundary terms 
vanish, due to the fact that the 
$\mathbf{q}$-integrand evaluated at $s=0$ is odd under $\mathbf{q} \mapsto - \mathbf{q}$. This yields
\begin{align}
    &i \nu \int \frac{\mathrm{d}^d   \mathbf{q}}{(2\pi)^d} \mathbf{q} \left\langle e^{i \mathbf{q}\cdot \mathbf{X}^{(0)}(t)}V_{-\mathbf{q}}\left[\phi_{\mathbf{q}}^{(1)}(t)  + i\mathbf{q}\cdot \mathbf{X}^{(1)}(t)  \phi_{\mathbf{q}}^{(0)}(t)\right] \right \rangle \n \\
    &=- \nu v \mathbf{\hat{e}_1} \int \frac{\mathrm{d}^d   \mathbf{q}}{(2 \pi )^d} \frac{q_1^2}{q^2 + r} \int_0^{\infty}\mathrm{d}s e^{-(\alpha_q - iq_1v) s - q^2({T}/{\kappa})\left(1 - e^{-\nu \kappa s }\right)}, 
\end{align}
where we assumed $\mathbf{v} = v \mathbf{\hat{e}_1}$. Using \cref{eq:2nd_order_part}, this leads to 
\begin{align}
     \left\langle\dot{\mathbf{X}}^{(2)}(t)\right\rangle + \nu \kappa \left\langle\mathbf{X}^{(2)}(t) \right\rangle = - \nu v \mathbf{\hat{e}_1} \int \frac{\mathrm{d}^d   \mathbf{q}}{(2 \pi )^d} \frac{q_1^2}{q^2 + r} \int_0^{\infty}\!\!\mathrm{d}s\, e^{-(\alpha_q - iq_1v) s - q^2({T}/{\kappa})\left(1 - e^{-\nu \kappa s }\right)}. 
\end{align}
This equation can be readily integrated by using the corresponding Green's function, leading to the expression for the time-independent shift reported in Eq.~\eqref{eq:shift}.
%of the main text. 

\textit{Sign of the shift. ---} In order to prove that the shift reported in \cref{eq:shift} is indeed in the direction opposite to the drag, it is sufficient to show that
\begin{equation}
    \int_0^{\infty}\!\!\mathrm{d}s\, \cos(q_1vs) \,\mathrm e^{- \alpha_q  s - (T/\kappa)q^2(1 - \mathrm e^{-\kappa \nu s})} > 0,
    \label{app:pos_shift_statement}
\end{equation}
because the imaginary part of the $\mathbf{q}$-integral in \cref{eq:shift} vanishes since the integrand is odd under $\mathbf{q} \mapsto -\mathbf{q}$. Let us denote
\begin{equation}
    f(s) = \mathrm e^{-a s -b (1-\mathrm e^{-c s})},
\end{equation}
with positive $a$, $b$, and $c$. 
By expanding the exponential in terms of its exponential argument, it is easy to prove that the function $f(s)$ can be expressed in the form
\begin{equation}
    f(s) = \int_0^\infty \dd{t} \mathrm e^{-st} \mathrm e^{-b} \sum_{n=0}^\infty \frac{b^n}{n!} \delta(t-a-n c),
\end{equation}
from which it follows that 
%
%\ag{I cut here some expressions...}
%
\begin{align}
    \int_0^\infty\!\! \dd{s}\, \cos(q_1 v s) f(s) 
    %&= \mathrm e^{-b} \sum_{n=0}^\infty \frac{b^n}{n!} \int_0^\infty \dd{t} \delta(t-a-n c) \left[ \int_0^\infty \dd{s} \mathrm e^{-st } \cos(q_1 v s) \right]  \n\\
    %&= \mathrm e^{-b} \sum_{n=0}^\infty \frac{b^n}{n!} \int_0^\infty \dd{t} \delta(t-a-n c) \frac{t}{t^2+(q_1 v)^2}\n\\
    &=\mathrm e^{-b} \sum_{n=0}^\infty \frac{b^n}{n!} \frac{a+n c}{(a+nc)^2+(q_1 v)^2} >0.
\end{align}
Choosing $a =\alpha_q$, $b = T/\kappa$, and $c = \kappa \nu$ proves the validity of Eq.~\eqref{app:pos_shift_statement}. Accordingly, the shift of the average particle position in the driven phase of the motion reported in \cref{eq:shift} is in the direction opposite to the drag.

\subsection{Recoil after the particle is released}
\label{app-recoil}

In deriving the expression of the particle position $\mathbf{X}$ after the particle is released from the trap ($t>0$), 
the solutions obtained for $t \leq 0$ serve as initial conditions. In particular, the field configuration at $t=0$ is the one given by Eq.~\eqref{field_t_neg}. When the particle and the field are decoupled (i.e., for $\lambda=0$), one obtains from 
\cref{eq:field_four} and \eqref{eq:part},
\begin{align}
    \mathbf{X}^{(0)}(t) &= \mathbf{X}^{(0)}(0) +  \mathbf{W}(t),
    \label{decoupled_part_t_pos}
     \\ 
     \phi^{(0)}_{\mathbf{q}}(t) &= \int_{-\infty}^t\mathrm{d}t'e^{-\alpha_q(t - t')} \eta_{\mathbf{q}}(t'),
    \label{decoupled_field_t_pos}
\end{align}
where $\mathbf{W}(t) = \int_{0}^{t}\mathrm{d}t' \bm{\xi}(t') $ is the Wiener process. This Gaussian process has a vanishing mean and a variance given by 
\begin{equation}
    \langle \mathrm{W}_i(t_1)\mathrm{W}_j(t_2)\rangle = d\nu T \min(t_1, t_2) \delta_{i,j}. 
\end{equation}
Again, using the results obtained in the decoupled case in Eqs.~\eqref{decoupled_part_t_pos} and \eqref{decoupled_field_t_pos}, we solve Eqs.~\eqref{eq:field_four} and  \eqref{eq:part} term-by-term according to the $\lambda$-expansion in Eqs.~\eqref{eq:particle_expansion} and \eqref{eq:field_expansion}. At the first order in $\lambda$, we obtain
    \begin{gather}
    \mathbf{X}^{(1)}(t) = \mathbf{X}^{(1)}(0) + i \nu \int \frac{\mathrm{d}^d   \mathbf{q}}{(2 \pi)^d}\mathbf{q}V_{-\mathbf{q}} e^{i\mathbf{q}\cdot \mathbf{X}^{(0)}(0)}\int_{0}^t \mathrm{d}t' e^{ i\mathbf{q}\cdot \mathbf{W}(t')} \phi_{\mathbf{q}}^{(0)}(t'),
     \\ 
     \phi^{(1)}_{\mathbf{q}}(t) = e^{-\alpha_q t} \phi^{(1)}_{\mathbf{q}}(0) + D_{\alpha}q^{\alpha}V_{\mathbf{q}}e^{-i\mathbf{q}\cdot \mathbf{X}^{(0)}(0)} \int_{0}^t \mathrm{d}t' e^{-\alpha_q(t - t') - i\mathbf{q}\cdot \mathbf{W}(t')}.
     \label{111}
\end{gather}
 We note that the first term in Eq.~(\ref{111}) is the contribution to the field from the relaxation over a time $t$ of its 
 steady-state configuration, characterized by the average reported in~\cref{eq:phi_NESS}. 
 To proceed, we recall Eq.~(\ref{eq:2nd_order_part})
and write
\begin{align}
     &i \nu \int \frac{\mathrm{d}^d   \mathbf{q}}{(2\pi)^d} \mathbf{q}  V_{-\mathbf{q}} \left\langle e^{i \mathbf{q}\cdot \mathbf{X}^{(0)}(t)}\phi_{\mathbf{q}}^{(1)}(t) \right \rangle =   i \nu \int \frac{\mathrm{d}^d   \mathbf{q}}{(2\pi)^d} \mathbf{q} V_{-\mathbf{q}}e^{-\alpha_q t}\left\langle e^{i \mathbf{q}\cdot\mathbf{W}(t)} \right \rangle \left\langle e^{i \mathbf{q}\cdot\mathbf{X}^{(0)}(0)} \phi_{\mathbf{q}}^{(1)}(0)\right \rangle \n \\ 
     &\qquad \qquad\qquad+ i \nu \int \frac{\mathrm{d}^d   \mathbf{q}}{(2\pi)^d} \mathbf{q} |V_{\mathbf{q}}|^2D_{\alpha} q^{\alpha}\int_0^{\infty}\mathrm{d} t' e^{-\alpha_q (t - t')} \left\langle e^{-i \mathbf{q}\cdot[\mathbf{W}(t) -\mathbf{W}(t')]} \right \rangle, 
     \label{222}
\end{align}
where we used the fact that the Wiener process performed by the particle after the trap is turned off is statistically independent from the dynamics at previous times.
We now note that
\begin{equation}
    \left\langle e^{i \mathbf{q}\cdot{[\mathbf{W}(t) - \mathbf{W}(t')]}} \right \rangle = e^{-q^2 \nu T |t - t'|},
\end{equation}
from which
\begin{equation}
    \left\langle e^{i \mathbf{q}\cdot{\mathbf{W}(t)}} \right \rangle = e^{-q^2 \nu Tt },
    \label{aaaaa}
\end{equation}
which is readily obtained by taking the limit $\kappa \to 0$ in Eq.~\eqref{eq:app_OU_SF}.
Note that the integrand in the second term of Eq.~(\ref{222}) is odd under $\mathbf{q} \mapsto - \mathbf{q}$, and therefore it vanishes. Accordingly, only the first term (stemming from the relaxation of the steady-state configuration) contributes to the force acting on the particle, as we mentioned in the main text above Eq.~(\ref{eq:field_relax_forcing}).  
We may now use the result obtained for $t \leq 0$ in \cref{field_t_neg} and write
\begin{align}
     &i \nu \int \frac{\mathrm{d}^d   \mathbf{q}}{(2\pi)^d} \left\langle\mathbf{q}e^{i \mathbf{q}\cdot \mathbf{X}^{(0)}(t)}V_{-\mathbf{q}}\phi_{\mathbf{q}}^{(1)}(t) \right \rangle   \\
     &= i \nu \int \frac{\mathrm{d}^d   \mathbf{q}}{(2\pi)^d}\mathbf{q} |V_{\mathbf{q}}|^2\frac{\alpha_q}{q^2 + r}e^{- (\alpha_q + q^2 \nu T)t} \int_{-\infty}^t\mathrm{d}t'e^{-\alpha_q(t - t') + i\mathbf{q}\cdot \mathbf{v}(t - t')-q^2({T}/{\kappa})\left[1 - e^{-\nu \kappa(t - t')} \right] }, \n
\end{align}
where we used $D_{\alpha}q^{\alpha} = \alpha_q/(q^2 + r)$ and \cref{{eq:app_OU_SF}}. 
Similarly, the second term in 
\cref{eq:2nd_order_part}
can be evaluated as
%
%\ag{I cut one line....}
%
\begin{align}
     &i \nu \int \frac{\mathrm{d}^d   \mathbf{q}}{(2\pi)^d} \mathbf{q} V_{-\mathbf{q}} \left\langle e^{i \mathbf{q}\cdot  \mathbf{X}^{(0)}(t)}i\mathbf{q}\cdot \mathbf{X}^{(1)}(t)  \phi_{\mathbf{q}}^{(0)}(t) \right \rangle \n\\
   % &=   i \nu \int \frac{\mathrm{d}^d   \mathbf{q}}{(2\pi)^d} \mathbf{q} V_{-\mathbf{q}} 
   %  \left\langle e^{i \mathbf{q}\cdot  \left[\mathbf{X}^{(0)}(0) + \mathcal{W}(t)\right]}i\mathbf{q}\cdot \mathbf{X}^{(1)}(0)  \phi_{\mathbf{q}}^{(0)}(t) \right \rangle   -  i\nu^2 \int \frac{\mathrm{d}^d   \mathbf{q}}{(2\pi)^d} \int \frac{\mathrm{d}^d   \mathbf{q}'}{(2 \pi)^d} \mathbf{q} V_{-\mathbf{q}}V_{-\mathbf{q'}} \n\\
     %&\qquad \qquad\times (\mathbf{q}\cdot\mathbf{q'}) \left\langle e^{i \mathbf{q}\cdot  \left[\mathbf{X}^{(0)}(0) + \mathbf{W}(t)\right]}e^{i\mathbf{q'}\cdot \mathbf{X}^{(0)}(0)}\int_{0}^t \mathrm{d}t' e^{ i\mathbf{q'}\cdot \mathbf{W}(t')} \phi_{\mathbf{q'}}^{(0)}(t')\phi_{\mathbf{q}}^{(0)}(t) \right \rangle   \n \\
     &=i \nu \int \frac{\mathrm{d}^d   \mathbf{q}}{(2\pi)^d} \mathbf{q} V_{-\mathbf{q}} \left\langle e^{i \mathbf{q} \cdot \mathbf{W}(t)} \right \rangle \left\langle  e^{i \mathbf{q}\cdot  \mathbf{X}^{(0)}(0) }i\mathbf{q}\cdot \mathbf{X}^{(1)}(0)  \phi_{\mathbf{q}}^{(0)}(t) \right \rangle  \n\\
     &\qquad \qquad +    i\nu^2 \int \frac{\mathrm{d}^d   \mathbf{q}}{(2\pi)^d}  \mathbf{q} |V_{\mathbf{q}}|^2q^2 \frac{T}{q^2 + r} \int_{0}^t \mathrm{d}t' e^{-(\alpha_q + q^2 \nu T)(t -t')}.
\end{align}
Again, the second term vanishes because its integrand is odd under $\mathbf{q} \to - \mathbf{q}$. We also note that 
\begin{equation}
    \phi_{\mathbf{q}}^{(0)}(t) = e^{-\alpha_q t} \phi_{\mathbf{q}}^{(0)}(0) + \int_0^t \dd{s} e^{-\alpha_q (t-s)}\eta_{\mathbf{q}}(s),
\end{equation}
%
%\ag{check the equation above, I modified it...}
%
where the second term is statistically independent 
of all other quantities evaluated at $t = 0$.
%
%\ag{comment on what happens to $\left\langle e^{i \mathbf{q} \cdot \mathbf{W}(t)} \right \rangle$ above?? Indicate where we discussed this....}
%
Hence, using \cref{aaaaa},  we can write
\begin{align}
    &i \nu \int \frac{\mathrm{d}^d   \mathbf{q}}{(2\pi)^d} \mathbf{q} V_{-\mathbf{q}} \left\langle e^{i \mathbf{q}\cdot  \mathbf{X}^{(0)}(t)}i\mathbf{q}\cdot \mathbf{X}^{(1)}(t)  \phi_{\mathbf{q}}^{(0)}(t) \right \rangle  =i \nu^2 \int \frac{\mathrm{d}^d   \mathbf{q}}{(2\pi)^d} \mathbf{q}|V_{\mathbf{q}}|^2q^2 \frac{T}{q^2 + r} e^{-(\alpha_q + q^2 \nu T)t}
    \n\\ 
    &\qquad\qquad\qquad\qquad\times
    \int_{-\infty}^t \mathrm{d}t'e^{-\alpha_q(t - t')-\nu \kappa(t - t') +  i \mathbf{q}\cdot \mathbf{v}\left(t - t'\right) - q^2({T}/{\kappa})\left(1- e^{-\nu \kappa (t - t')} \right)}. 
\end{align}
This allows us to cast the average of \cref{eq:2nd_order_part} in the form
%
%\ag{Why do we talk here abut time derivative of the particle position? In the previous expressions there was no time derivative... Which expression should I look at??}
%
\begin{align}
\left \langle\dot{\mathbf{X}}^{(2)}(t)\right \rangle  =- \nu v \mathbf{\hat{e}_1} \int \frac{\mathrm{d}^d   \mathbf{q}}{(2 \pi )^d} \frac{q_1^2}{q^2 + r} e^{-(\alpha_q + q^2 \nu T)t}\int_0^{\infty}\mathrm{d}s e^{-(\alpha_q - iq_1v) s - q^2({T}/{\kappa})\left(1 - e^{-\nu \kappa s }\right)} .
\end{align}
Simple integration over $t$ yields the final result for the position of the particle, reported in Eq.~(\ref{eq:recoil_full}) in the main text.

\section{Analysis of the recoil}

In this Appendix, we further characterize the recoil discussed in \cref{sec:recoil,sec:asympt}, by deriving explicit expressions and evaluating asymptotic behaviors. 

\subsection{Small-\texorpdfstring{$v$}{v} behavior of \texorpdfstring{$\mathcal{F}^{A/B}$}{F(A/B)}}
\label{app:small_v}

In this Appendix, we derive the small-$v$ behavior of the scaling function $\mathcal{F}^{A/B}$ defined in Eqs.~\eqref{zeroT_rec_scale_def} and
\eqref{zeroT_rec_scale}, characterizing the recoil in the noiseless limit ($T \to 0$). The non-trivial dependence of its scaling properties on the spatial dimensionality $d$,  reported in Tab.~\ref{tab:small_v_scale}, stems from the fact that the limit $v \to 0$ may not commute with the limits $\xi \to \infty$ and $t \to \infty$.  

In this and the following appendices, we 
drop the 
symbol tilde 
%`` $\tilde{\cdot}$ " 
on top of variables  $\tilde \xi, \tilde t$  and $\tilde{v}$ defined in Eqs.~\eqref{eq:dimless_var} and \eqref{eq:Weissenberg}, understanding that 
henceforth
$t$, $\xi$, and $v$ are dimensionless. 
%
%\ag{why don't we use the same notation as in the main text?}
%
In  order to obtain the asymptotic small-$v$ behavior of the scaling functions defined in Eq.~(\ref{zeroT_rec_scale}), we use the Mellin transform. For a function $f(u)$ defined for $u\ge 0$, the Mellin transform $ \mathcal{M}\{f \}(s) $ is defined as\cite{Davies_book}
\begin{equation}
    \mathcal{M}\{f \}(s) = \int_0^{\infty}\mathrm{d}u\, u^{s-1} f(u),
\end{equation}
where $s\in \mathbb{C}$ within a certain domain $\mathcal D$ of analiticity,  with the inverse
\begin{equation}
    f(u) = \int_{c - i \infty}^{c + i \infty} \frac{\mathrm{d}s}{2 \pi i } u^{-s} \mathcal{M}\{f \}(s),
    \label{mellin_inverse}
\end{equation}
where the integral runs within $\mathcal D$.
%
%\ag{I have added some specifications above, check that you agree....}
%
We perform the transform (from the variable $v$ to the variable $s$) of the denominator of the function on the right-hand side of Eq.~(\ref{zeroT_rec_scale}), obtaining 
\begin{equation}
    \mathcal{M}\left\{\frac{1}{\alpha_q^2  + q_1^2v^2}\right\}(s)  = \frac{\pi }{2}\frac{1}{\sin(\pi s/2)} \frac{1}{\alpha_q^2 }\left( \frac{\alpha_q}{q_1} \right)^s, \quad \mbox{with} \quad 0<\Re(s)<2,
\end{equation}
where 
%in the adopted short-hand notation 
$\alpha_q = q^{\alpha}(q^2 + 1/\xi^2)$ (see \cref{eq:alpha_q}).
Note that $1/\sin(\pi s/2)$
has simple poles when $s$ equals an even integer, in particular for $s=0$ and $s=2$ which are at the boundary of the domain of analiticity of the transform above. According to Eq.~(\ref{mellin_inverse}),  these latter two poles determine the behavior $\sim v^0$ for $v \ll 1$, and  $\sim v^{-2}$ for large $v$ of the scaling function defined in \cref{zeroT_rec_scale}.
%
%\ag{Which function?}
%
The Fourier transform of the Gaussian interaction kernel introduced in \cref{eq:Gauss_kernel}, given by $\hat{V}_{\mathbf{q}}=e^{-q^2/2}$,  
%
%\ag{refer to the formula in the text where this was written first...}
%
regularizes the large-$q$ behavior of the integrand in Eq.~(\ref{zeroT_rec_scale}).
Accordingly, any non-trivial analytic behavior can arise solely from possible infrared divergences ($q\to 0$) in the $v \to 0$ limit. 
This means that, for any finite $t$, in order to study the small-$q$ behavior of Eq.~(\ref{zeroT_rec_scale}), one can use  $1 - e^{-\alpha_q t}  \approx \alpha_q t$ for critical model A, and for model B (for any value of $\xi$). 
The integrand of \cref{zeroT_rec_scale}
%
%\ag{of what?}
%
in the case of non-critical model A turns out to exhibit the same small-$q$ behavior both for finite and infinite $t$ --- this follows from the fact that its modes are characterized by a finite relaxation time at low wavevectors, see Eq.~(\ref{eq:alpha_q}). 
Also, the angular integral
in \cref{zeroT_rec_scale}
is inconsequential in the discussion of possible infrared divergences present for  $v \to 0$. In fact, note that in 
\begin{equation}
\mathcal{F}^{A/B}\left( \frac{1}{\xi}, 
t, v \right) \xrightarrow[v \to 0]{} \int\frac{\mathrm{d}^d  \mathbf{q}}{(2 \pi)^d}\frac{q_1^2}{q^2 + 1/\xi^2}|\hat{V}_{\mathbf{q}}|^2\frac{1 - e^{-tq^{\alpha}\left(q^2 +1/\xi^2\right)}}{\left[q^{\alpha}\left(q^2 +1/\xi^2\right)\right]^2},
\label{app_zeroT_rec_scale}
\end{equation}
the angular integral 
only involves the term $q_1^2$ at the numerator
and it always yields a finite result.  Hence, we focus on the radial integral in \cref{zeroT_rec_scale}, i.e.,
\begin{equation}
    \mathcal{M}\left\{ \mathcal{F}^{A/B}\left(1/\xi, t \right)\right\}(s)  \sim  \frac{t}{\sin(\pi s/2)}\int_0^{\infty} \!\!\mathrm{d}q \,q^{d-1} e^{-q^2} q^{\alpha} \left( \frac{q_1}{\alpha_q}\right)^{2-s} ,\qquad t < +\infty,
    \label{eq:app_mellin_t_fin}
\end{equation}
where we used the expression for $\alpha_q$ in \cref{eq:alpha_q},
%
%\ag{refer here to some other equation where we have written this... it is certainly not the first time...}
%
and
\begin{equation}
    \mathcal{M}\left\{ \mathcal{F}^{A/B}\left(1/\xi, t \right)\right\}(s)  \sim  \frac{1}{\sin(\pi s/2)}\int_0^{\infty}\!\! \mathrm{d}q \,q^{d-1} \frac{e^{-q^2}}{q^2 + 1/\xi^2} \left( \frac{q_1}{\alpha_q}\right)^{2-s} ,\quad\mbox{for}\quad t \to \infty.
    \label{eq:app_mellin_t_inf}
\end{equation}
We recall that, for small-$q$, one has $\alpha_q\sim q^{\alpha}$ for finite $\xi$, and $\alpha_q\sim q^{\alpha +2}$ for $\xi \to \infty$. This yields the small-$q$ asymptotics of the integrands reported in Tab.~\ref{tab:Gammas}.
%%
%%
%%
%%%%%%%%%%%%%%%%%%%%%%%%%%%%%%%%%%%%%%%%%%%%%%%%%%
%%%%%%%%%%%%%%%%%%%%%%%%%%%%%%%%%%%%%%%%%%%%%%%%%%
\begin{table}
\begin{center}
\begin{tabular}{ | m{4em} | m{4cm} m{4cm} | } 
  \hline
   $\alpha = 0$ 
   & $\xi < +\infty$
  & $\xi \to \infty$ \\ 
  \hline
  $t < + \infty$ & $\sim q^{d-s + 1}$ & $\sim q^{d+s-3}$ \\ 
  
  $t \to \infty$ & $\sim q^{d-s + 1}$ & $\sim q^{d+s-5}$ \\ 
  \hline
\end{tabular}

\vspace{.2cm}
\begin{tabular}{ | m{4em} | m{4cm} m{4cm} | } 
  \hline
   $\alpha = 2$ 
   & $\xi < +\infty$
  & $\xi \to \infty$ \\ 
  \hline
  $t < + \infty$ & $\sim q^{d+s-1}$ & $\sim q^{d+3s-5}$ \\ 
  
  $t \to \infty$ & $\sim q^{d+s-3}$ & $\sim q^{d+3s-9}$ \\ 
  \hline
\end{tabular}
\end{center}
\caption{Behavior for small $\mathbf{q}$ of the integrands of the Mellin transforms (from variable $v$ to $s$) of the scaling function in Eqs.~\eqref{eq:app_mellin_t_fin} and \eqref{eq:app_mellin_t_inf}. The poles that emerge after integration of these functions, according to the inverse Mellin transform in Eq.~(\ref{mellin_inverse}), correspond to the small-$v$ behavior of the scaling function reported  in Tab.~\ref{tab:small_v_scale}.
%of the main text. 
%
}
\label{tab:Gammas}
\end{table} 
%%%%%%%%%%%%%%%%%%%%%%%%%%%%%%%%%%%%%%%%%%%%%%%%%%
%%%%%%%%%%%%%%%%%%%%%%%%%%%%%%%%%%%%%%%%%%%%%%%%%%
%%
%%
%%
Note that, whenever the exponent governing the small-$q$ behavior of the integrand is equal to $-1$, the corresponding function has a simple pole. This observation, in conjunction with the expression for the inverse Mellin transform in Eq.~(\ref{mellin_inverse}), leads to the small-$v$ behavior of the scaling function reported in Tab.~\ref{tab:small_v_scale}.
%in the main text.

\subsection{Long-\texorpdfstring{$t$}{t} and large-\texorpdfstring{$\xi$}{xi} behavior of \texorpdfstring{$\mathcal{F}^{A/B}$}{F(A/B)}}
 \label{app_lin_scale}

In this Appendix, we derive the asymptotic behavior of the scaling function of the recoil defined in Eqs.~\eqref{zeroT_rec_scale_def} and \eqref{zeroT_rec_scale} at long times,
found 
%in the main text 
in Eqs.~\eqref{eq:crit_recoil_asympt}, \eqref{eq:subcrit_A_recoil_asympt}, and \eqref{eq:subcrit_B_recoil_asympt},  and for large correlation lengths, after taking $t \to \infty$, as reported in Eq.~\eqref{eq:range_asympt}.  In the derivation presented below, a finite value of $v$ is assumed in order to avoid the issues discussed in \ref{app:small_v}.

Here we adopt the short-hand notation introduced in \ref{app:small_v} and we focus on the case of finite $v$.
Assuming that $\mathcal{R}^A(0,v) = \mathcal{F}^A\left( 0, t\to \infty, v\right)$ exists, we can write
  \begin{align}    
 \mathcal{R}^A\left(0, v\right) -  \mathcal{F}^A\left( 0, t, v\right) &=   \int\frac{\mathrm{d}^d  \mathbf{q}}{(2 \pi)^d}\frac{q_1^2}{q^2}|\hat{V}_{\mathbf{q}}|^2\frac{ e^{-q^2t}}{q^4 + q_1^2 v^2} =  t^{1-d/2}\int\frac{\mathrm{d}^d  \mathbf{q}}{(2 \pi)^d}\frac{q_1^2}{q^2}|\hat{V}_{\frac{\mathbf{{q}}}{\sqrt{t}}}|^2\frac{ e^{-q^2}}{q^4/t + q_1^2 v^2} \n\\
 &\underset{t \gg 1}{\sim} \frac{t^{(2-d)/2}}{v^2}|\hat{V}_0|^2\int\frac{\mathrm{d}^d  \mathbf{q}}{(2 \pi)^d}\frac{e^{-q^2}}{q^2} = \frac{t^{(d-2)/2}}{v^2} \frac{2 |\hat{V}_0|^2}{(4 \pi)^{d/2}(d-2)}.
 \label{app_rec_crit_t_scale}
 \end{align}
 From the scaling $\sim t^{(2-d)/2}$ in the second line of this expression,
 it follows that $\mathcal{F}^{A}\left(0, t\to \infty, v\right)$ is finite only for $d \geq 2$. Moreover, the $q$-integral in the last line is divergent in the vicinity of $q\sim 0$ for $d\leq 2$.
 The cases of $d=1$ or $2$ are studied separately
 in \ref{app:rec_full},
 for the specific choice of the Gaussian interaction kernel introduced in \cref{eq:Gauss_kernel}.
 Note that the right-hand side of Eq.~(\ref{app_rec_crit_t_scale}) is proportional to $ 1/v^2$ for any dimensionality. Moreover, the factor ${2 |\hat{V}_0|^2}/[{(4 \pi)^{d/2}(d-2)}]$ in front of Eq.~(\ref{app_rec_crit_t_scale}) changes sign as a function of the dimensionality $d$. In particular, it is positive for $d >2$ and negative for $d<2$. Accordingly, the analytic continuation to $d<2$ describes the asymptotic long-time divergence of the critical recoil which was verified by a direct calculation of the critical scaling function in the case of model A in $d = 1$ dimensions, see Eq.~(\ref{app_crit_rec_1}).  
 
Similarly, in the case of model B at criticality one finds
  \begin{align}
 \mathcal{R}^B\left(0, v\right) -  \mathcal{F}^B\left( 0, t, v\right) &=   \int\frac{\mathrm{d}^d  \mathbf{q}}{(2 \pi)^d}\frac{q_1^2}{q^2}|\hat{V}_{\mathbf{q}}|^2\frac{ e^{-q^4t}}{q^8 + q_1^2 v^2}  \label{app_rec_crit_t_scale_B}\\
 &=  t^{1/2-d/4}\int\frac{\mathrm{d}^d  \mathbf{q}}{(2 \pi)^d}\frac{q_1^2}{q^2}|\hat{V}_{\frac{\mathbf{{q}}}{t^{1/4}}}|^2\frac{ e^{-q^4}}{q^8/t^{3/2} + q_1^2v^2}  \n \\  
 &\underset{t \gg 1}{\sim} \frac{t^{(2-d)/4}}{v^2} |\hat{V}_0|^2\int\frac{\mathrm{d}^d  \mathbf{q}}{(2 \pi)^d}\frac{e^{-q^4}}{q^2} = \frac{t^{(2-d)/4}}{v^2}  \frac{4 \pi^{1/2 }|\hat{V}_0|^2}{(8 \pi)^{d/2}\Gamma(d/4)(d-2)},
\n
 \end{align}
 where we assumed $d>2$ in order to evaluate the integral in the last line. Similarly 
 to the case of
 model A discussed above, the analytic continuation of the expression for $d<2$ yields a correct description of the long-time divergence of the recoil in the case of critical model B at $d=1$. 
In order to simplify the notation, we introduce 
\begin{equation}
    \mathcal{C}_{\alpha}(d) = \begin{cases}
   2/[(4 \pi)^{d/2}(2-d)], \quad &\mbox{for} \quad \alpha =0, \\ 
        4 \pi^{1/2 }/[(8 \pi)^{d/2}\Gamma(d/4)(2-d)], \quad &\mbox{for} \quad \alpha = 2,
    \end{cases}
    \label{app:c_alpha}
\end{equation}
as we did in Eq.~(\ref{eq:crit_recoil_asympt}). Both $\mathcal{C}_0(d)$ and $\mathcal{C}_2(d)$ have simple poles at $d=2$ with the same residue $-{1}/({2 \pi})$. At $d=2$, instead of power laws, one observes a logarithmic growth in time of the scaling function $\mathcal{F}^{A/B}$ in  \cref{zeroT_rec_scale} for large times (see, c.f., \cref{app_crit_rec_2}). 
%
%\ag{In what? What do you mean?}

For non-critical model A,  the long-$t$ behavior is obtained as
\begin{align}    
\mathcal{R}^A\left(1/\xi, v\right) -  \mathcal{F}^A\left( 1/\xi, t, v\right) &=   \int\frac{\mathrm{d}^d  \mathbf{q}}{(2 \pi)^d}\frac{q_1^2}{q^2 + 1/\xi^2}|\hat{V}_{\mathbf{q}}|^2\frac{ e^{-(q^2 + 1/\xi^2)t}}{(q^2 + 1/\xi^2)^2 + q_1^2 v^2}  \label{app_rec_crit_t_A_subcrit_scale}\\
&=\frac{e^{-t/\xi^2}}{t^{d/2 + 1}}
\int\frac{\mathrm{d}^d \mathbf{q}}{(2 \pi )^d} \frac{q_1^2}{ q^2/t + 1/\xi^2}|\hat{V}_{ \mathbf{q}/\sqrt{t}}| \frac{e^{-q^2 }}{\left(q^2/t + 1/\xi^2\right)^2 + q_1^2 v^2/t} \n\\
&\underset{t \gg 1}{\sim}  \frac{\xi^6 e^{-t/\xi^2}}{t^{d/2 + 1}} |\hat{V}_0|^2 \int\frac{\mathrm{d}^d \mathbf{q}}{(2 \pi )^d} q_1^2e^{-q^2} =  \frac{|\hat{V}_0|^2}{2 (4 \pi)^{d/2}}\frac{\xi^6 e^{-t/\xi^2}}{t^{d/2 + 1}} ,\n
 \end{align}
which 
%is 
corresponds to the expression
reported in Eq.~(\ref{eq:subcrit_A_recoil_asympt}).
%in the main text. 
Finally, for non-critical model B we find
 \begin{align}    
 \mathcal{R}^B\left(1/\xi, v\right) -  \mathcal{F}^B\left( 1/\xi, t, v\right) &=    \int\frac{\mathrm{d}^d  \mathbf{q}}{(2 \pi)^d}\frac{q_1^2}{q^2 + 1/\xi^2}|\hat{V}_{\mathbf{q}}|^2\frac{ e^{-q^2(q^2 + 1/\xi^2)t}}{q^4(q^2 + 1/\xi^2)^2 + q_1^2 v^2} \n\\
 &=\frac{\xi^{d + 2}}{t^{d/2}}
\int\frac{\mathrm{d}^d \mathbf{q}}{(2 \pi )^d} \frac{q_1^2}{q^2 \xi^4/t + 1}|\hat{V}_{\mathbf{q} \frac{\xi}{\sqrt{t}}}| \frac{e^{-q^2 -q^4\xi^4/t}}{q^4 \xi^2/t\left(q^2\xi^2/t + 1/\xi^2\right)^2 + q_1^2 v^2} \n \\
 &\underset{t \gg 1}{\sim} 
\frac{\xi^{d + 2}}{t^{d/2}v^2}|\hat{V}_0|^2 \int\frac{\mathrm{d}^d \mathbf{q}}{(2 \pi )^d} e^{-q^2} =  \frac{|\hat{V}_0|^2}{(4 \pi)^{d/2}}\frac{\xi^{d + 2}}{ v^2 t^{d/2}},
 \label{app_rec_crit_t_B_subcrit_scale}
 \end{align}
 which is the expression reported in Eq.~(\ref{eq:subcrit_B_recoil_asympt}). 
 %
% \ag{check if some of the intermediate expressions above can be omitted...}
 %

%    
% QUIQUIQUIQUIQUIQUIQUI
%

For a finite correlation length $\xi$, the limit $\mathcal{F}^{A/B}\left(1/\xi, t\to \infty , v\right)$ exists for any dimensionality. We 
can thus
 write
  \begin{align}
 \mathcal{R}^{A/B}\left( 1/\xi, v\right) &= \int\frac{\mathrm{d}^d  \mathbf{q}}{(2\pi)^d}\frac{q_1^2}{q^2 + 1/\xi^2}\frac{|\hat{V}_{\mathbf{q}}|^2}{[q^{\alpha}(q^2 + 1/\xi^2)]^2 + q_1^2v^2} \n \\ 
 &= \xi^{2-d}\int \frac{\mathrm{d}^d   \mathbf{q}}{(2 \pi)^d}\frac{q_1^2}{1 + q^2} \frac{|\hat{V}_{\frac{\mathbf{q}}{\xi^2}}|^2}{[q^{\alpha}(1 + q^2)]^2/\xi^{2(\alpha + 1)} + q_1^2v^2} \n \\ 
 &\underset{\xi \to \infty}{\sim}
 \frac{\xi^{2-d}}{v^2} |\hat{V}_0|^2  \int \frac{\mathrm{d}^d   \mathbf{q}}{(2 \pi)^d}\frac{1}{1 + q^2}  =  |\hat{V}_0|^2\frac{ \Gamma(1 -d/2)}{(4 \pi)^{d/2}} \frac{\xi^{2-d}}{v^2},
 \end{align}
leading to  Eq.~(\ref{eq:range_asympt}).  Notice that the integral in the last line yields a finite result for $d<2$, while it exhibits divergences at large $q$ for $d \geq 2$.

\subsection{Exact expressions for the critical recoil with model A dynamics}
\label{app:rec_full}
%\textbf{[Check this again in the morning!!!]}

It turns out that the integral in the expressions \eqref{zeroT_rec_scale_def} and  \eqref{zeroT_rec_scale}  for the scaling function of the recoil can be calculated exactly in the case of model A with $\xi \to \infty$ when the Gaussian interaction kernel (see \cref{eq:Gauss_kernel}) is considered. Here we report the results of such an integration, and discuss its main features. 

For $d=1$, the analytic expression for the scaling function of the critical recoil turns out to be
\begin{equation}    
\mathcal{F}^A\left( 0, t, v\right) =  \frac{1}{v^2} \frac{\sqrt{1 +t} - 1
}{\sqrt{\pi}} + \frac{1}{2v^3} \left[ e^{ v^2(1 + t)}\erfc\left(v\sqrt{1 +t}\right) - e^{ v^2}\erfc\left(v\right) \right],
\label{app_crit_rec_1}
\end{equation}
where $ \erfc(x) = 1 - (2/\sqrt{\pi})\int_0^x \mathrm{d}u e^{-u^2}$ is the complementary error function. Accordingly, the long-time behavior
of Eq.~\eqref{app_crit_rec_1}
is given by
 \begin{equation}
\mathcal{F}^A\left(0, t, v\right)  \simeq \frac{\sqrt{t}}{\sqrt{\pi}v^2} - \left[ \frac{1}{\sqrt{\pi}v^2} + \frac{1}{2v^3}e^{v^2}\erfc{\left(v\right)}\right], \quad\mbox{for}\quad t \gg 1,
 \label{app_crit_rec_1_as}
 \end{equation}
 where we used the fact that $e^{x^2}\erfc(x)\simeq {1}/{(\sqrt{\pi}x)}$ for large $x$.
 Accordingly, upon increasing $t$,
 this expression diverges like $\sqrt{t}$. Note that the prefactor $1/\sqrt{\pi}$ corresponds to $\mathcal{C}_{0}(1)$ given by Eq.~(\ref{app:c_alpha}). Moreover, expanding the expression in the square parenthesis in Eq.~(\ref{app_crit_rec_1}) around $v = 0$  shows that the critical scaling function (see Eq.~(\ref{zeroT_rec_scale}) with $R/\xi \to 0$) behaves like $v^{-1}$ for $v \ll 1$. This is in agreement with the expression reported in Tab.~\ref{tab:small_v_scale}.

 Next, in $d=2$, we find
\begin{equation}    
\mathcal{F}^A\left( 0, t, v \right) =   \frac{\ln(1 + t)
}{4 \pi v^2} + \frac{1}{4 \pi v^2}\left[e^{v^2(1+t)/2}K_0\left(\frac{v^2(1+t)}{2}\right) 
 -  e^{v^2/2}K_0\left(\frac{v^2}{2}\right)\right],
 \label{app_crit_rec_2}
\end{equation}
where $K_0$ is the modified Bessel function of the second kind. Since $e^xK_0(x) \simeq \sqrt{\pi/(2x)}$, the recoil exhibits a logarithmic divergence at long times. Also, the critical scaling function in Eq.~(\ref{zeroT_rec_scale}) behaves asymptotically as $\propto \ln v$ for small $v$, in agreement with Tab.~\ref{tab:small_v_scale}.

Finally, in $d=3$
\begin{align}    
\mathcal{F}^A\left( 0, t, v\right) =& \,   \frac{1}{8 \pi
v^3}\left[\frac{1 - e^{v^2(1 + t)}\erfc\left(v\sqrt{1 + t}\right)}{1 + t} - \left(1 - e^{v^2}\erfc\left(v\right) \right) \right] \n\\
&\qquad\qquad-\frac{1
}{4\pi^{3/2}v^2}\left(\frac{1}{\sqrt{1 + t}}-1\right) ,
 \label{app_crit_rec_3}
\end{align}    
hence
 \begin{equation}
\mathcal{F}^A\left( 0, t, v\right)  \sim \frac{1
}{4\pi^{3/2}v^2}\left[1   - \frac{\sqrt{\pi}}{2v}\left(1  - e^{v^2}\erfc{\left( v\right)} \right)\right] - \frac{1}{4\pi^{3/2}v^2}\frac{1}{\sqrt{t}},
 \quad \mbox{for}\quad t \gg 1,
 \label{app_crit_rec_3_as}
 \end{equation}
where we used the behavior of $\erfc(x)$ at large $x$ reported after Eq.~\eqref{app_crit_rec_1_as}. Accordingly,  the scaling function in Eq.~(\ref{zeroT_rec_scale}) --- and consequently the recoil ---  approaches asymptotically a finite value from below as $t^{-1/2}$. Again, the prefactor $-{1}/({4 \pi ^{3/2}})$ agrees with the expression of $\mathcal{C}_{0}(3)$ given in Eq.~(\ref{app:c_alpha}). Expanding  Eq.~(\ref{app_crit_rec_3}) around $v=0$ for finite $t$ shows that the critical scaling function in Eq.~\eqref{app_crit_rec_3} exhibits the asymptotic $\sim v^0$ behavior 
%\dav{do you really mean ``1'' or rather $v^0$?} 
for small $v$, in agreement with Tab.~\ref{tab:small_v_scale}. 
However, taking first the limit $t \to \infty$ in \cref{{app_crit_rec_3}} and then expanding around $v = 0$ yields an asymptotic $\sim v^{-1}$ behavior, again in agreement with the scaling reported in Tab.~\ref{tab:small_v_scale}.

\section{Field configuration with a driven particle}
\label{app:crit_tail}

In this Appendix, we analyze the steady-state configuration of the average field in the reference frame co-moving with the harmonic potential. The Fourier transform of this field in the noiseless limit is given by  \cref{eq:Z_NESS_T0} in the main text. In particular, we consider separately the cases of model A and model B dynamics for the field, in \ref{app:crit_tail_A} and \ref{app:crit_tail_B}, respectively, focusing on the behavior far from the particle. 

\subsection{Model A}
\label{app:crit_tail_A}

In order to determine the behavior of the field profile $\varphi^{(\mathrm{ss})}(y)$ in the noiseless limit $T=0$, given by Eq.~(\ref{eq:Z_NESS_T0}), away from the particle, i.e., for $|y|\gg R$ , one can conveniently approximate the interaction potential $V(\mathbf{x})$ with that of a point-like particle, i.e., $V(\mathbf{x}) \propto  \delta^{(d)}(\mathbf{x})$. 
%whose Fourier transform is $V_{\mathbf{q}} =1$. 
We thus evaluate in Cartesian coordinates
\begin{align}
    \varphi^{(\mathrm{ss})}(y) &= \lambda D_0 \int \frac{\mathrm{d}^d \mathbf{q}} {(2 \pi)^d}\frac{e^{i q_1 y}}{D_0q^2 - iq_1v} =  \lambda D_0 \int_0^{\infty}\mathrm{d} u \int \frac{\mathrm{d}^d \mathbf{q}} {(2 \pi)^d}e^{-u(D_0q^2 - iq_1 v) + iq_1 v}   
    \n \\ 
    %&= \lambda D_0 \int_0^{\infty}\mathrm{d}u \left[\int \frac{\mathrm{d}q}{2 \pi} e^{-D_0q^2 u}\right]^{d-1}\int \frac{\mathrm{d q_1}}{2 \pi}e^{-D_0q_1^2 u + iq_1 (uv + y)} \n \\
    %&=   \lambda D_0 \int_0^{\infty} \frac{\mathrm{d}u}{(4 \pi D_0 u)^{d/2}} e^{ - \frac{(uv + y)^2}{4Du}}   
    %=   \lambda D_0 e^{-\frac{y v}{2D_0}}\int_0^{\infty}\frac{\mathrm{d}u}{(4 \pi D_0 u)^{d/2}} e^{ - \left(u \frac{v^2}{4D_0} + u^{-1} \frac{y^2}{4D_0}\right)}  \n \\ 
    &=\frac{2}{(4 \pi)^{d/2}} \lambda \left( \frac{v}{D_0 |y|}\right)^{(d - 2)/2}e^{-yv/(2D_0)}K_{(d -2)/2}\left(\frac{v |y|}{2 D_0} \right).
\end{align}
%
%\ag{Above I cut some intermediate steps...}
%
Using the asymptotic behavior of the modified Bessel function of the second type
\begin{equation}
    K_n(x) \simeq e^{-x}\sqrt{\frac{\pi}{2x}},  \quad  \mbox{for} \quad x \gg 1 ,
\end{equation}
mentioned previously in the case of $n=0$ (in the footnote~\footref{foot} on~\Cpageref{foot}), %\ag{I think this was already written somewhere in the manuscript... refer to it...}
one obtains
\begin{equation}
    \varphi^{(\mathrm{ss})}(y) \simeq \frac{\lambda}{(4 \pi |y|)^{\frac{d-1}{2}}} \left(\frac{v}{D_0} \right)^{(d-3)/2} e^{-v(y + |y|)/(2 D_0)}, \quad\mbox{for}\quad |y| \gg 1.
\end{equation}
Note that 
%for 
the left tail (corresponding to $y<0$) of the steady-state configuration of the field exhibits an algebraic decay $\sim |y|^{-(d-1)/2}$. Moreover, the right tail (for $y>0$) is characterized by the lengthscale reported in Eq.~(\ref{eq:lplus_crit}), which characterizes the exponential decay of the profile upon moving away from the particle.

\subsection{Model B}
\label{app:crit_tail_B}

For model B, it is convenient to introduce the dimensionless variable $\tilde r = 2^{2/3}r D_2^{2/3}/(3v^{2/3})$, which is connected to the lengthscale introduced in \cref{eq:xi-v} 
%of the main text 
%Eq.~(\ref{eq:xi_v_def}) 
via
\begin{equation}
    \tilde{r} = \frac{2^{2/3}}{3}\left(\frac{l_{v,2}}{\xi}\right)^2.
\end{equation}
For $\tilde{r}<1$, we define
\begin{equation}
\beta\left(\tilde{r}\right) = \left( 1 + \sqrt{1 - \tilde{r}^3} \right)^{1/3},
\end{equation}
and evaluate the inverse Fourier transform of Eq.~(\ref{eq:Z_NESS_T0}) which, for $y<0$, is given by
\begin{equation}
    \varphi^{(\mathrm{ss})}(y) = \frac{\lambda (2D_2)^{1/3}}{3v^{1/3}}\frac{(\beta + \tilde{r} \beta^{-1})e^{-|y|(v/2D_2)^{1/3} \left( \beta + \tilde{r}\beta^{-1} \right)}}{\beta^2 + \tilde{r} + \tilde{r}^2 \beta^{-2}},
\end{equation}
while, for $y>0$, by
\begin{align}
&\varphi^{(\mathrm{ss})}(y) = \frac{\lambda (2D_2)^{1/3}}{3v^{1/3}}\frac{(\beta + \tilde{r} \beta^{-1})e^{-y(v/2D_2)^{1/3}\left( \beta + \tilde{r}\beta^{-1} \right)/2}}{\left(\beta^2 + \tilde{r} + \tilde{r}^2 \beta^{-2}\right)} \times \label{eq:app_shadow_b_2} \\
&\left[\cos \left(\frac{\sqrt{3}}{2} y(v/2D_2)^{1/3}\left( \beta - \tilde{r}\beta^{-1} \right)\right) - \sqrt{3}\frac{\beta^2 + \tilde{r}^2\beta^{-2}}{\beta^2 - \tilde{r}^2\beta^{-2}} \sin \left(\frac{\sqrt{3}}{2} y(v/2D_2)^{1/3}\left( \beta - \tilde{r}\beta^{-1} \right) \right) \right], \n
\end{align}
where the argument $\tilde r$ has been omitted from $\beta (\tilde r)$ for brevity.  Note that, as a function of $y$, the term in the second line is characterized by spatial oscillations with a typical wavevector $k$
%
%\ag{check notation with the main text...}
%
which vanishes as $\sim \sqrt{1 - \tilde{r}}$ when $\tilde{r} \to 1$ from below. 
For $\tilde{r}>1$, we additionally define
\begin{equation}
\theta\left(\tilde{r}\right) = \frac{1}{3}\arctan\left(\left(\tilde{r}^3 - 1\right)^{-1/2} \right).
\label{eq:app_shadow_b_1}
\end{equation}
Then, for $d=1$ and $\tilde r>1$, a standard evaluation of the inverse Fourier transform of Eq.~(\ref{eq:Z_NESS_T0}) by the means of a complex integration yields
\begin{equation}
    \varphi^{(\mathrm{ss})}(y) = \frac{\lambda \xi}{2 \sqrt{3}}\frac{\cos\left(\frac{\pi}{6}  - \theta\right)}{ \cos \theta \sin\left(\frac{\pi}{6}  + \theta\right)}e^{-\frac{|y|}{\xi} \frac{2}{\sqrt{3}}\cos\left( \frac{\pi}{6} - \theta \right)}, \qquad \text{for } y<0,
    \label{aaa}
         \end{equation}
%
%\ag{the combinations of $\cos$ and $\sin$ here can be written as $\cos$ of something + phase... check and simplify this expression and the one which follows.}
%
and
 %\begin{align}
  %  \varphi^{(\mathrm{ss})}(y) =& \, \frac{\lambda\xi(\cos \theta - \frac{1}{\sqrt{3}}\sin \theta)}{2 \cos \theta(\cos \theta - \sqrt{3} \sin \theta)}e^{-\frac{y}{\xi}\left(\cos \theta - \frac{1}{\sqrt{3}}\sin \theta\right)}  \n\\ 
 %   &- \frac{2\lambda\xi\sin \theta}{\sqrt{3}(\cos \theta + \sqrt{3} \sin \theta) (\cos \theta - \sqrt{3} \sin \theta)}e^{-\frac{y}{\xi} \frac{2}{\sqrt{3}}\sin \theta}, \qquad \text{for }  y>0,
%\label{eq:app_phi_b_ypl_noosc}
%\end{align}
%asdsadsa
\begin{align}
    \varphi^{(\mathrm{ss})}(y) =& \, \frac{\lambda \xi}{2 \sqrt{3}}\frac{\cos\left( \frac{\pi}{6} + \theta \right)}{ \cos \theta \sin \left(\frac{\pi}{6} - \theta\right)}e^{-\frac{y}{\xi}\frac{2}{\sqrt{3}}\cos\left( \frac{\pi}{6} + \theta \right)}  \n\\ 
    &- \frac{ \lambda \xi}{2\sqrt{3}}\frac{\sin \theta}{\sin \left( \frac{\pi}{6} + \theta\right) \sin \left( \frac{\pi}{6} - \theta\right)}e^{-\frac{y}{\xi} \frac{2}{\sqrt{3}}\sin \theta}, \qquad \text{for }  y>0,
\label{eq:app_phi_b_ypl_noosc}
\end{align}
where we dropped the argument $\tilde{r}$ in  $\theta$ for brevity. 
Note that the $y$-integral over $\mathbb{R}$ vanishes, as expected because of the local conservation of the field. 
%
%\ag{right?}
%
In passing, we note that \rev{taking} the limit $v \to 0$ ($\tilde{r} \to \infty$) in \cref{aaa} and \cref{eq:app_phi_b_ypl_noosc} \rev{recovers the} %\rev{\st{which is the}} 
field configuration at equilibrium. %\dav{???}  
In addition, taking $\xi \to \infty$ yields the result reported in Eq.~(\ref{eq:shadow_tail_B}) of the main text.
Importantly, the qualitative behavior of the steady-state field configuration for $y>0$ in the case of model B dynamics, which is either characterized by a superposition of two exponential functions as in Eq.~(\ref{eq:app_phi_b_ypl_noosc}), or  by damped oscillations as in Eq.~(\ref{eq:app_shadow_b_2}), holds for any spatial dimensionality $d$ of the system. In fact, these two qualitatively distinct behaviors are determined by the value of the parameter $\tilde{r}$ and follow from the analysis of the poles of Eq.~(\ref{eq:Z_NESS_T0}), which holds for any $d$.

\section*{References}
\bibliography{refs}
\end{document}